\documentclass{emulateapj}

\usepackage{graphicx}
\usepackage[normalem]{ulem}                        
\usepackage{txfonts}
\usepackage{amssymb}
\usepackage{enumerate}
\usepackage[usenames,dvipsnames]{color}           
\newcommand{\BE}{\begin{equation}}
\newcommand{\EE}{\end{equation}}
\newcommand{\BA}{\begin{eqnarray}}
\newcommand{\EA}{\end{eqnarray}}

\renewcommand{\vec}[1]{{\mathbf #1}}
\newcommand{\deriv}[2]{\frac{{\mathrm d} #1}{{\mathrm d} #2}}

\newcommand{\surf}{ {\mathcal S} }

\newcommand{\bb}{\vec B}

\newcommand{\xx}{ \vec x}

\newcommand{\uu}{ \vec u}
\newcommand{\vv}{ \vec v}

\newcommand{\gth}{G_{\theta}}
\newcommand{\gph}{G_{\Phi}}

\newcommand{\kev}[1]{{\color{black}{#1}} \color{black}}  

\newcommand{\gr}[1]{{\color{BrickRed}{\sout{#1}}}}

\newcommand{\eq}[1]{Equation~(\ref{eq:#1})} 
\newcommand{\eqs}[2]{Equations~(\ref{eq:#1}) and (\ref{eq:#2})} 
\newcommand{\eqss}[2]{Equations~(\ref{eq:#1}) -- (\ref{eq:#2})} 
\newcommand{\sect}[1]{Section~\ref{sec:#1}}

\newcommand{\app}[1]{Appendix~\ref{app:#1}}

\newcommand{\tab}[1]{Table~\ref{tab:#1}}

\newcommand{\fig}[1]{Figure~\ref{fig:#1}}

\newcommand{\cf}{\textit{cf.} }
\newcommand{\eg}{\textit{e.g.}, }
\newcommand{\ie}{\textit{i.e.}, }

\shorttitle{Studying the transfer of magnetic helicity in solar active regions}
\shortauthors{Dalmasse et al.}

\begin{document} 

   \title{Studying the transfer of magnetic helicity in solar active regions\\
   with the connectivity-based helicity flux density method}
   
   \author{
	K. Dalmasse$^{1}$,
	\'E. Pariat$^{2}$,
	G. Valori$^{3}$,
	J. Jing$^{4}$,
	and P. D\'emoulin$^{2}$
	}
		
   \affil{$1$ CISL/HAO, National Center for Atmospheric Research, 
                 P.O. Box 3000, Boulder, CO 80307-3000, USA
                 }
	\email{dalmasse@ucar.edu}
   \affil{$2$ LESIA, Observatoire de Paris, PSL Research University, CNRS, 
   		Sorbonne Universit\'e, UPMC Univ. Paris 06, Univ. Paris Diderot, 
		Sorbonne Paris cit\'e, 
		F-92190 Meudon, France
                 }
   \affil{$3$ Mullard Space Science Laboratory, University College London, 
   		Holmbury St. Mary, Dorking, Surrey, RH5 6NT, UK
                 }
   \affil{$4$ Space Weather Research Laboratory, Center for Solar Terrestrial Research, 
   		New Jersey Institute of Technology, 323 Martin Luther King Blvd., Newark, 
		NJ 07102-1982, USA
                 }

   \begin{abstract}
In the solar corona, magnetic helicity slowly and continuously accumulates in response to plasma 
flows tangential to the photosphere and magnetic flux emergence through it. Analyzing this transfer 
of magnetic helicity is key for identifying its role in the dynamics of active regions (ARs). 
The connectivity-based helicity flux density method was recently developed for studying the 2D and 3D 
transfer of magnetic helicity in ARs. The method takes into account the 3D nature of magnetic helicity 
by explicitly using knowledge of the magnetic field connectivity, which allows it to faithfully track 
the photospheric flux of magnetic helicity. Because the magnetic field is not measured in the solar 
corona, modeled 3D solutions obtained from force-free magnetic field extrapolations must be used 
to derive the magnetic connectivity. Different extrapolation methods can lead to markedly different 
3D magnetic field connectivities, thus questioning the reliability of the connectivity-based approach 
in observational applications. We address these concerns by applying this method to the 
\kev{isolated and internally} 
complex AR 11158 with different magnetic field extrapolation models. We show that the connectivity-based 
calculations are robust to different extrapolation methods, in particular with regards to identifying 
regions of opposite magnetic helicity flux. We conclude that the connectivity-based approach can be 
reliably used in observational analyses and is a promising tool for studying the transfer of magnetic 
helicity in ARs and relate it to their flaring activity.
   \end{abstract}   

   \keywords{magnetic fields - Sun: photosphere - Sun: corona - Sun: flares
               }

%

\section{Introduction} \label{sec:S-Introduction}

%
%
Magnetic helicity is a signed scalar quantity that measures the three-dimensional complexity of 
a magnetic field in a volume \citep[\eg][]{Finn85}. \cite{Moffatt69} and \cite{Berger84} showed 
that magnetic helicity has a well-defined geometrical interpretation in terms of the entanglement, 
or braiding, of magnetic field lines. Magnetic helicity thus generalizes more local properties such 
as magnetic twist and shear.

%
%
The emergence of twisted/sheared magnetic fields from the convection zone into the solar corona 
\citep[\eg][]{Leka96,MorenoInsertis97,Longcope00,Demoulin02,Green02,Georgoulis09,Pevtsov12,Poisson15}, 
and the stressing of the coronal magnetic field by plasma flows along the photosphere 
\citep[\eg][]{vanBallegooijen89,Klimchuk92,Chae01b,Moon02,Liu12,Zhang12,Vemareddy15}, slowly 
and continuously build up magnetic helicity in the solar atmosphere. Because of its conservation 
property in highly conducting plasmas \citep[\eg][]{Woltjer58,Taylor74,Taylor86,Berger84b,Pariat15}, 
magnetic helicity is thus believed to be a fundamental component for understanding the dynamics 
of the coronal magnetic field \citep[\eg][]{Zhang06,Zhang08,Kazachenko12,Tziotziou12,Romano14}.

%
%
Magnetic helicity is hence at the heart of several MHD theories of coronal processes including, 
but not limited to, coronal heating through the relaxation of braided magnetic fields 
\citep[\eg][]{Heyvaerts84,Russell15,Yeates15}, the formation of filament channels through the inverse 
cascade of magnetic helicity \citep[\eg][]{Antiochos13,Knizhnik15}, the existence of CMEs as the mean 
for the Sun to expel its magnetic helicity excess \citep[\eg][]{Rust94,Low96}, and the production of very 
high-energy flares via magnetic helicity annihilation \citep{Linton01,Kusano04}. Recently, \cite{Pariat17} 
even showed that specific quantities derived from  magnetic helicity have a strong potential for greatly 
improving the prediction of solar eruptions.

%
%
Methods to estimate magnetic helicity in the solar context are reviewed by \cite{Valori16}. Among these 
methods, analyzing the temporal evolution of the helicity flux through the photosphere provides valuable 
information about the helicity content of ARs and is one of the means for better understanding the role 
of magnetic helicity in their dynamics \citep[see review by \eg][and references therein]{Demoulin09}. 
When an AR is followed from the beginning of its emergence, the temporal integration of the photospheric 
helicity flux gives an estimate of its coronal helicity 
\citep[\eg][]{chae01,Kusano03,Mandrini04,Jeong07,Labonte07,Yang09,Guo13}. On the other hand, 
the photospheric distribution of helicity flux during the early stages of AR formation reflects 
the sub-photospheric distribution of magnetic helicity in the associated emerging magnetic field 
\citep[\eg][]{Kusano03,Chae04,Yamamoto05,Pariat06,Jing12,Park13,Vemareddy17}. 
This, in turn, gives constraints on the processes generating the magnetic field in the solar interior 
\citep[\eg][]{Kusano02,Pariat07}. Later on during the lifetime of an AR, the distribution of the helicity flux 
allows to track where magnetic helicity is being locally accumulated in response to additional magnetic 
flux emergence and photospheric flows \citep[\eg][]{Chandra10,Vemareddy12}.

%
%
Studying the distribution of the helicity flux in ARs is not straightforward because it requires the use of 
a surface density of a quantity which is inherently 3D and not local. While magnetic helicity density 
per unit volume is an unphysical quantity, \cite{Russell15} recently showed that it is possible to construct 
and study a magnetic helicity density per unit surface from the recent developments of 
\cite{Yeates11,Yeates13,Yeates14}. Previously, \cite{Pariat05} had shown that it was possible to define 
a useful proxy of surface density of helicity flux by explicitly expressing magnetic helicity in terms of 
magnetic field lines linkage. Such an approach is achieved by including the connectivity of magnetic 
field lines in the definition of the total helicity flux, leading to the construction of a so-called 
{\it connectivity-based} surface density of helicity flux (further details are provided in \sect{S-Hflux-densities}).

%
%
\cite{Dalmasse14} recently developed a method for the practical computation of the connectivity-based 
helicity flux density to be used in observational studies. Using analytical case-studies and numerical 
simulations, they showed that the connectivity-based calculations provide a reliable and faithful mapping 
of the helicity flux. In particular, the method is successful in revealing real mixed signals of helicity flux 
in magnetic configurations, as well as in relating the local transfer of magnetic helicity with the location 
of regions favorable to magnetic reconnection. The former makes the method particularly interesting 
for testing the very high-energy flare model of \cite{Kusano04} in observational surveys of solar ARs, 
while the latter provides a new way for analyzing the role of magnetic helicity accumulation in flaring 
activity.

%
%
For analytical models and numerical MHD simulations, the 3D magnetic field is known in the entire 
volume of the modeled solar atmosphere and can be readily used to integrate magnetic field lines. 
In observational studies, however, polarimetric measurements in the corona are not as numerous 
and routinely made as the photospheric and chromospheric ones. And as the latters, coronal 
polarimetric measurements are also 2D and, thus, cannot lead to magnetic field data in the full coronal 
volume without the use of some 3D modeling. On top of this, their inversion into magnetic field data is a very 
challenging task \citep[\eg][and references therein]{Rachmeler12,Kramar14,Plowman14,Dalmasse16,Gibson16}. 
Hence, one must rely on the approximate 3D solution of, \eg nonlinear force-free field (NLFFF) models 
\citep[\eg][]{Wheatland00,Wiegelmann04,Amari06,Valori07,Inoue12,Malanushenko12} 
to extrapolate the coronal magnetic field from the photospheric maps of the magnetic field (vector 
magnetograms). Unfortunately, different methods and assumptions can lead to markedly different 
3D magnetic field solutions. These strong differences between reconstructed magnetic fields affect 
all subsequently derived quantities. For instance, \citet{DeRosa09,DeRosa15} reported variations 
between extrapolation methods that can reach up to $30\%$ in free magnetic energy and $200 \%$ 
in magnetic helicity.

%
%
The analyses of \citet{DeRosa09,DeRosa15} raise concerns about the reliability and 
relevance of the connectivity-based helicity flux density approach in observational applications. 
In this paper, we address these concerns by applying the connectivity-based method to observations of 
\kev{an AR} 
with different magnetic field extrapolation models and implementations. 
\kev{The selected AR is internally complex but externally simple (i.e., no neighboring large-flux systems).}  
The connectivity-based helicity flux density method is reviewed in \sect{S-Method}. \sect{S-Observations} 
describes the dataset and the approach taken to estimate uncertainties in the helicity flux intensity. 
The magnetic field extrapolations are discussed in \sect{S-Extrapolations}. \sect{S-Results} presents 
the results of our analysis. A discussion and interpretation of our results is provided in \sect{S-Discussion}. 
Our conclusions are summarized in \sect{S-Conclusion}.


\section{Method} \label{sec:S-Method}

\subsection{Magnetic Helicity Flux Densities} \label{sec:S-Hflux-densities}

%
%
Under ideal conditions, the transfer of magnetic helicity through the photosphere, $\surf$, 
and into the solar atmosphere is \citep[\eg][]{Demoulin02b,Pariat05}
\BE
	\label{eq:Eq-Hflux}
	\deriv{H}{t} = \int_{\surf} \gth (\xx) \ \mathrm{d} \surf \,.
\EE
$\gth (\xx)$ is a surface-density of magnetic helicity flux defined as
\BE
	\label{eq:Eq-Gtheta}
	\gth (\xx) = - \frac{B_n  (\xx)}{2 \pi} \int_{\surf'} \deriv{\theta ( \xx - \xx' )}{t} B_n  (\xx') \ \mathrm{d} \surf'  \,.
\EE
where
\BE
	\deriv{\theta ( \xx - \xx' )}{t} = \frac{\left( \left( \xx - \xx' \right) \times \left( \uu - \uu' \right) \right) |_n}{|\xx - \xx'|^2}  \,,
\EE
is the relative rotation rate (or relative angular velocity) between pairs of photospheric magnetic 
polarities located at $\xx$ and $\xx'$ and moving on the photospheric plane $\surf$ with flux-transport 
velocity $\uu = \uu(\xx)$ and $\uu' = \uu(\xx')$. The flux-transport velocity $\uu$ is \citep{Demoulin03}
\BE
	\label{eq:Eq-FTV}
	\uu = \vv_t - \frac{v_n}{B_n} \bb_t   \,.
\EE
where subscript ``$t$'' and ``$n$'' respectively denote the tangential and normal components 
of photospheric vector fields, $\vv$ and $\bb$ are the photospheric plasma velocity and magnetic 
fields.

%
%
\eqss{Eq-Hflux}{Eq-FTV}  show that the total flux of magnetic helicity through the photosphere, 
$\surf$, can be expressed as the summation of the net rotation of all pairs of photospheric 
elementary magnetic polarities around each other, weighted by their magnetic flux. It further 
shows that, at any given time, $t$, the total flux of magnetic helicity can be solely derived from 
photospheric quantities, \ie from a timeseries of vector magnetograms from which the flux-transport 
velocity can also be derived \citep{Schuck08}.

%
%
The surface-density of helicity flux, $\gth$, measures the variation of magnetic helicity in an AR 
only from the relative motions of its photospheric magnetic polarities. However, we recall that 
magnetic helicity describes the global linkage of magnetic field lines in the volume. Therefore, 
what effectively modifies the magnetic helicity of an AR is the relative re-orientation of magnetic 
field lines, \ie the variation of their mutual helicity, in response to the motions of their photospheric 
footpoints. Such a global, 3D nature of magnetic helicity and its variation is not taken into account 
in the definition of $\gth$. As a consequence, $\gth$ has a tendency to misrepresent 
the local variation of magnetic helicity in ARs by hiding the subtle effects of the mutual helicity 
variation between magnetic field lines, and by overestimating the local helicity flux when an AR 
is associated with opposite helicity fluxes \citep[\eg][]{Dalmasse14}.

%
%
\cite{Pariat05} showed that it is possible to remedy this issue by explicitly re-arranging those terms 
in \eq{Eq-Hflux} that are related to the connectivity of elementary magnetic flux tubes. That allows 
to express the total flux of magnetic helicity in terms of the sum of the magnetic helicity variation 
of each individual elementary flux tube of the magnetic field, such that
\BE
	\label{eq:Eq-Hflux-co}
	\deriv{H}{t} = \int_{\Phi} \deriv{h_{\Phi}}{t} \bigg|_c \ \mathrm{d} \Phi_{c}   \,,
\EE
where $h_{\Phi}$ is the magnetic helicity of the elementary magnetic flux tube, $c$, and 
$\mathrm{d} \Phi_{c}$ its elementary magnetic flux. $h_{\Phi}$ is the magnetic helicity density 
and describes how any elementary flux tube is linked/winded around all other elementary flux tubes 
\citep[\eg][]{Berger88,Aly14,Yeates14}. It can be shown that
\BA
	\label{eq:Eq-hflux-dens-1}
	\deriv{h_{\Phi}}{t} \bigg|_c & = & \frac{\gth (\xx_{c_+})}{|B_n (\xx_{c_+})|} + \frac{\gth (\xx_{c_-})}{|B_n (\xx_{c_-})|}     \\
	\label{eq:Eq-hflux-dens-2}
	 & = & \dot{\Theta}_{B} (\xx_{c_+}) - \dot{\Theta}_{B} (\xx_{c_-})   \,,
\EA
where $\xx_{c_+}$ (resp. $\xx_{c_-}$) is the positive (resp. negative), photospheric, magnetic 
polarity of the elementary magnetic flux tube $c$ 
\citep[see Figure 1 of][for a generic sketch of elementary flux tubes and their photospheric connectivity]{Dalmasse14}, 
and 
\BE
	\label{eq:Eq-Omega}
	\dot{\Theta}_{B} (\xx) = - \frac{1}{2 \pi} \int_{\surf'} \deriv{\theta ( \xx - \xx' )}{t} B_n  (\xx') \ \mathrm{d} \surf'  \,.
\EE
Note that the advantage of using \eq{Eq-hflux-dens-2} over \eq{Eq-hflux-dens-1} is to avoid artificial 
numerical singularities when $B_n (\xx)$ is very small.

Then, \cite{Pariat05} defined a new surface density of helicity flux by redistributing 
$\mathrm{d} h_{\Phi} / \mathrm{d} t$ at each photospheric footpoint of the elementary magnetic 
flux tube, $c$, to map the photospheric flux of magnetic helicity
\BE
	\label{eq:Eq-Gphi}
	\gph (\xx_{c_{\pm}}) = \frac{1}{2} \deriv{h_{\Phi}}{t} \bigg|_c \left|B_n (\xx_{c_{\pm}}) \right|   \,.
\EE
Note that the factor 1/2 in \eq{Eq-Gphi} assumes that both photospheric footpoints of an elementary 
flux tube contribute equally to the variation of its magnetic helicity 
\citep[a more general case is described in][]{Pariat05}. Note also that the calculations of 
$\mathrm{d} h_{\Phi} / \mathrm{d} t$ and $\gph$ require one significant additional information 
as compared with $\gth$, \ie the connectivity of the magnetic field.

Finally, we stress here that both $\gth$ and $\gph$ are \textit{instantaneous} estimations of injected 
helicity at particular location on the photosphere, and not the total helicity content in the associated 
field line.


\subsection{Numerical Method} \label{sec:S-Numerical-Method}

\cite{Dalmasse14} introduced a method to compute the connectivity-based helicity flux density 
proxy, $\gph$. Using various analytical case studies and numerical MHD simulations, they showed 
that $\gph$ properly tracks the site(s) of magnetic helicity variations.

%
%
Their method is based on field line integration to derive the magnetic connectivity required 
to compute $\gph$. For observational studies, routine magnetic field measurements are mostly 
realized at the photospheric level (\cf \sect{S-Introduction}). We thus perform force-free field 
extrapolations to obtain the coronal magnetic field of the studied AR (see \sect{S-Extrapolations}) 
and compute the photospheric distribution of magnetic helicity flux from \eqs{Eq-hflux-dens-2}{Eq-Gphi}. 

%
%
Each pixel of the photospheric vector magnetogram is identified as the cross-section of an elementary 
magnetic flux tube with the photosphere. Each of these elementary flux tubes is associated with one 
magnetic field line that is integrated to obtain the connectivity. For any closed magnetic field line, we thus 
obtain a pair ($ \xx_{c_{+}} ; \xx_{c_{-}}$) of photospheric footpoints, where $ \xx_{c_{+}}$ is the positive 
magnetic polarity of the elementary flux tube and $ \xx_{c_{-}}$ is its negative magnetic polarity at the 
photosphere ($z = 0$). We then introduce a slight modification to the method proposed in \cite{Dalmasse14}. 
Instead of computing $\gph$ from \eq{Eq-hflux-dens-1}, which can lead to artificial numerical singularities 
when $B_n (\xx)$ is very small, we compute $\mathrm{d} h_{\Phi} / \mathrm{d} t$ and $\gph$ from 
\eqs{Eq-hflux-dens-2}{Eq-Gphi}. Hence, once the conjugate footpoint of a field line is found, we compute 
$\left( \dot{\Theta}_{B} (\xx_{c_+}) \right.$ ; $\left. \dot{\Theta}_{B} (\xx_{c_-}) \right)$ using bilinear 
interpolation. Finally, field lines for which the conjugate footpoint reaches the top or lateral boundaries 
are treated as open field lines and both $\mathrm{d} h_{\Phi} / \mathrm{d} t$ and $\gph$ are simply set 
to zero. This is a second modification to the method of \cite{Dalmasse14} justified by the fact that, 
in this paper, we focus on comparing $\mathrm{d} h_{\Phi} / \mathrm{d} t$ and $\gph$ computed using 
different magnetic field extrapolations, \ie two quantities that are only defined for closed field lines.


\subsection{Metrics for Validation and Quantification of $\gph$ Maps} \label{sec:S-Metrics}

%
%
In the connectivity-based helicity flux density approach, half the total helicity flux computed from 
$\gth$ over the ensemble of positive magnetic polarities is redistributed over the ensemble of 
negative magnetic polarities, and vice-versa. For this redistribution to be perfect, the total magnetic 
flux summed over the magnetic polarities where $\gph$ is computed must vanish. If, for instance, 
the magnetic flux from the negative magnetic polarities is smaller than that of the positive ones, 
then a fraction of the total helicity flux computed from $\gth$ over the positive magnetic polarities 
is not redistributed in the negative ones. This means that part of the total helicity flux computed 
with $\gth$ for the entire magnetic configuration is missing from that computed with $\gph$. 
We thus introduce a first metric to validate $\gph$ maps, \ie the percentage of magnetic flux 
imbalance, $\tau_{\Phi_{\mathrm{imb.}}}$, for the closed magnetic flux where $\gph$ is computed
\BE
	\label{eq:Eq-Tau}
	\tau_{\Phi_{\mathrm{imb.}}} = \frac{ \int_{\surf_{\mathrm{cl.}}} B_n (\xx) \ \mathrm{d} \surf }{\min \left( \int_{\surf_{\mathrm{cl.}} \left(B_n > 0 \right)} B_n (\xx) \ \mathrm{d} \surf \ ; \ \left| \int_{\surf_{\mathrm{cl.}} \left(B_n < 0 \right)} B_n (\xx) \ \mathrm{d} \surf \right| \right)}   \,.
\EE
where $\surf_{\Phi_{\mathrm{cl.}}}$ is the part of the photosphere associated with the closed magnetic 
flux, $\Phi_{\mathrm{cl.}}$. 

%
%
By definition, the fluxes measured by $\gph$ are simply a redistribution of the fluxes measured 
by $\gth$ at both footpoints of each elementary magnetic flux tube in the closed magnetic field. 
Thus, the intensity of the magnetic helicity flux in one pixel can be different for $\gph$ and $\gth$. 
However, the total flux of magnetic helicity integrated over the closed magnetic flux from $\gph$ 
must be the same as that integrated from $\gth$ because \eq{Eq-Hflux-co} is equal to \eq{Eq-Hflux} 
for the closed magnetic field. The second metric we defined for the validation of $\gph$ maps is 
thus
\BE
	\label{eq:Eq-Cs-closed}	
	C_{\surf_{\Phi_{\mathrm{cl.}}}} = 
	   \frac{ \int_{\surf_{\Phi_{\mathrm{cl.}}}} \left( \gph (\xx) - \gth (\xx) \right) \ \mathrm{d} \surf }
	        { \int_{\surf_{\Phi_{\mathrm{cl.}}}} \gth (\xx) \ \mathrm{d} \surf }  \,.
\EE
In theory, $C_{\surf_{\Phi_{\mathrm{cl.}}}}$ should be strictly equal to 0. 
In practice, however, its value depends on departures of $\tau_{\Phi_{\mathrm{imb.}}}$ from 0. 
Accuracy and validation of the $\gph$ map requires both that $\tau_{\Phi_{\mathrm{imb.}}}$ and 
$C_{\surf_{\Phi_{\mathrm{cl.}}}}$ be close to 0.

%
%
Finally, we define two quantification indices to compare the signal intensity between $\gph$ and $\gth$ 
within the closed magnetic field
\BA
	\label{eq:Eq-Cp}
	C_{+} & = & \frac{ \int_{\surf_{\mathrm{cl.}} \left( \gth > 0 \right)} \gth (\xx) \ \mathrm{d} \surf }{ \int_{\surf_{\mathrm{cl.}} \left( \gph > 0 \right)} \gph (\xx) \ \mathrm{d} \surf }  \,,   \\
	\label{eq:Eq-Cm}
	C_{-} & = & \frac{ \int_{\surf_{\mathrm{cl.}} \left( \gth < 0 \right)} \gth (\xx) \ \mathrm{d} \surf }{ \int_{\surf_{\mathrm{cl.}} \left( \gph < 0 \right)} \gph (\xx) \ \mathrm{d} \surf }  \,.
\EA
$C_{+}$ and $C_{-}$ respectively compare the total positive and total negative magnetic helicity 
fluxes derived from $\gth$ with those derived from $\gph$ in the closed magnetic flux. \cite{Pariat05} 
and \cite{Dalmasse14} showed that the helicity flux density proxy $\gth$ hides the true local helicity flux 
and tends to exhibit larger and spurious helicity flux intensities as compared with the $\gph$ proxy when 
opposite helicity fluxes are present in an AR. $C_{+}$ and $C_{-}$ allow to quantify the intensity of 
the spurious signals in $\gth$, thus providing an idea of the global improvement of the $\gph$ maps 
relatively to the $\gth$ ones. In particular, large departures of $C_\pm$ from 1 are indicative of 
strong spurious signals in $\gth$.


\section{Observations and error analysis} \label{sec:S-Observations}

\subsection{Data} \label{sec:S-Data}

  \begin{figure}
   \centerline{\includegraphics[width=0.46\textwidth,clip=]{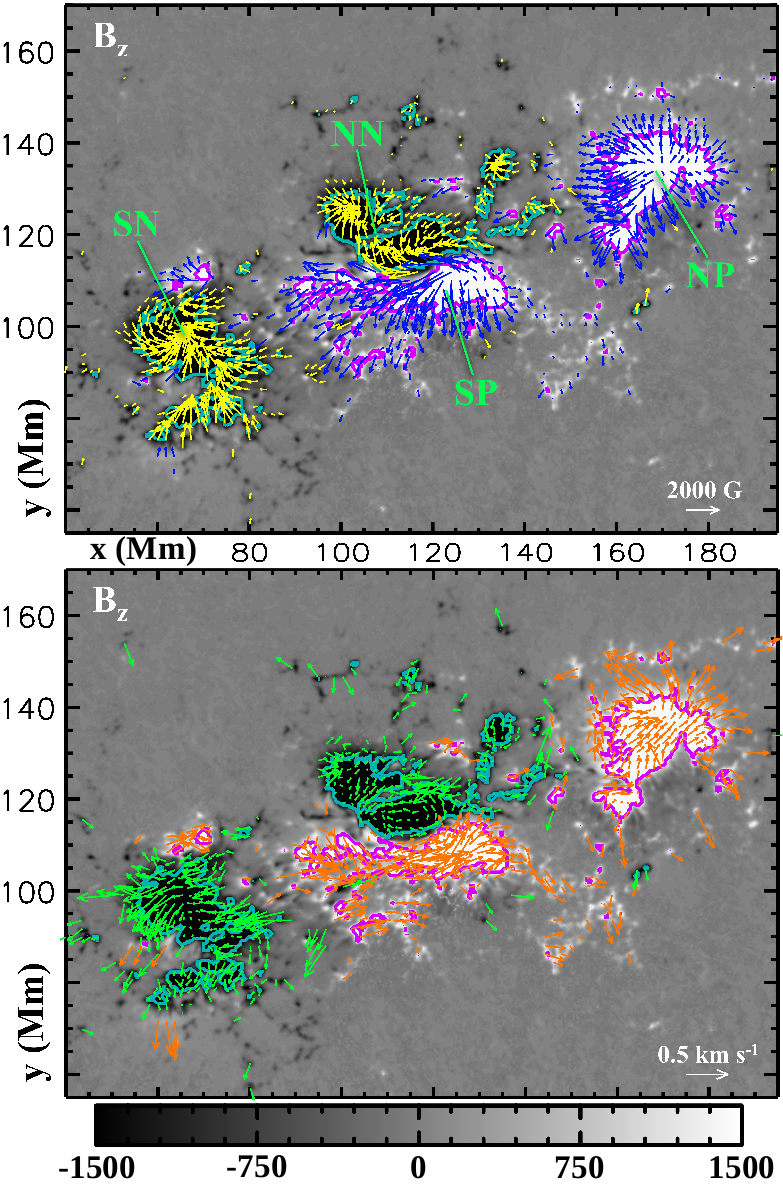}
              }
   \caption{{\bf Top:} SDO/HMI photospheric vector-magnetogram at $\sim$ 06:28 UT. The gray scale displays the vertical component, $B_z$ (in Gauss), while the yellow/blue arrows show the transverse component of the magnetic field. {\bf Bottom:} vertical magnetic field overplotted with the flux transport velocity field (green/orange arrows). Pink and cyan solid lines show $B_z = \pm 500 \ \mathrm{Gauss}$ isocontours.}
              \label{fig:Fig-BFTV}
   \end{figure} 

%
%
We test the robustness of the connectivity-based helicity flux density method against different 
magnetic field extrapolation models of the \kev{internally} complex AR 11158. This AR appeared 
on the solar disk at the heliographic coordinates S19 E42 on 2011 February 10. The AR was the result 
of fast and strong magnetic flux emergence that produced two large-scale bipoles, a northern 
and a southern one, in close proximity \citep[see \fig{Fig-BFTV}; \eg][]{Schrijver11}. 
The complex, quadrupolar magnetic field of this AR produced several C-/M-/X-class flares and 
CMEs during its on-disk passage \citep[\eg][]{Toriumi14}. A large fraction of this flaring activity 
was associated with the collision between the negative magnetic polarity of the northern bipole, 
NN, and the positive magnetic polarity of the southern bipole, SP, which led to a strong and 
continuous shearing of their polarity inversion line. More details on the configuration, evolution, 
and flaring activity of the AR can be found in \eg \cite{Sun12}, \cite{Jiang12}, \cite{Vemareddy12b} 
and \cite{Inoue13}.

%
%
The coronal models of AR 11158 are computed using vector magnetograms taken by the Helioseismic 
and Magnetic Imager \citep[HMI, \eg][]{Schou12} onboard the {\it Solar Dynamics Observatory} 
\citep[SDO, \eg][]{Pesnell12}. {\it SDO}/HMI provides full-disk vector magnetograms of the Sun 
with a pixel size of 0.5$''$. For the purpose of this paper, we re-use part of the data from \cite{Dalmasse13} 
who had applied the connectivity-based approach to AR 11158 with an NLFFF extrapolation without 
addressing the possible dependency of the results on the choice of extrapolation method. In particular, 
vector magnetograms at 06:22 UT and 06:34 UT on 2011 February 14 from the HMI-SHARP data series, 
HARP number 377 \citep{Hoeksema14} are re-used.

These two vector magnetograms are used to derive the photospheric flux transport velocity field 
with the differential affine velocity estimator for vector magnetograms \citep[DAVE4VM;][]{Schuck08}, 
using a window size of 19 pixels as suggested by \cite{Liu13}. The computed flux transport velocity 
field effectively represents an instantaneous flux transport velocity field at $\sim$ 06:28 UT. The two 
vector magnetograms taken at 06:22 UT and 06:34 UT are averaged to construct an instantaneous 
vector magnetogram associated with the flux transport velocity field at $\sim$ 06:28 UT. The constructed 
vector magnetic field is then used both to compute $\gth$ and as the photospheric boundary condition 
for the different FFF extrapolations presented \sect{S-Extrapolations}. The vector magnetic field and 
flux transport velocity field at $\sim$ 06:28 UT are shown in \fig{Fig-BFTV}.

  \begin{figure*}
   \centerline{\includegraphics[width=0.98\textwidth,clip=]{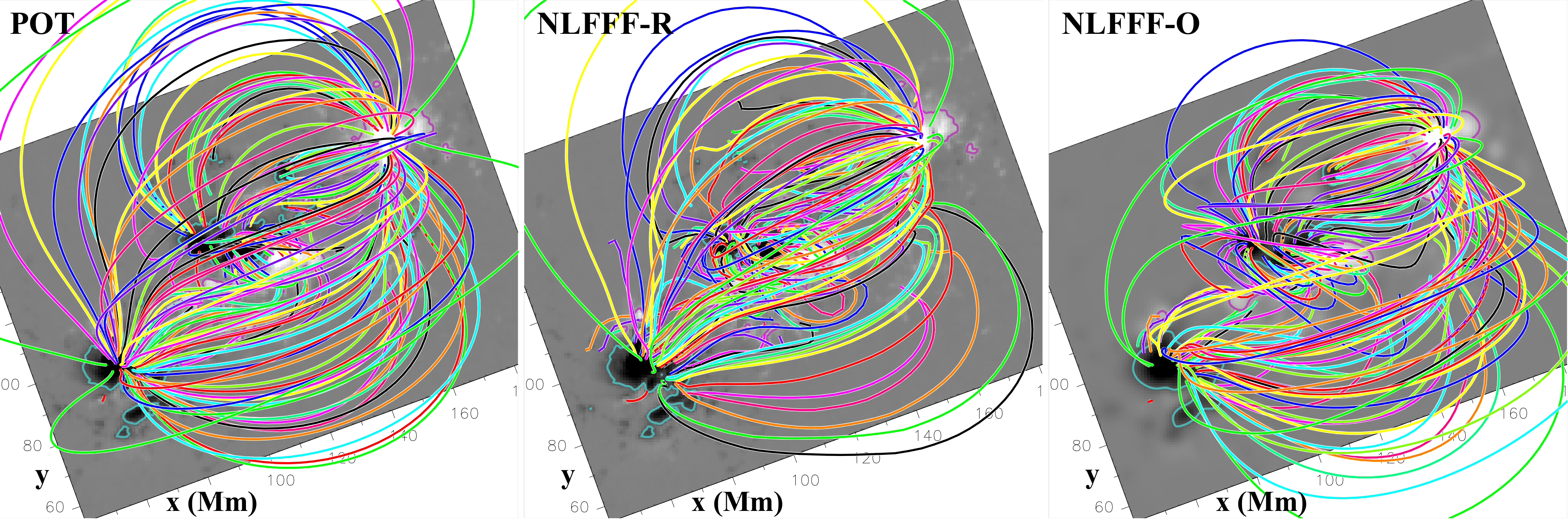}
              }
   \caption{Selected \kev{closed} field lines of the 3D extrapolated magnetic field for the potential field (labelled POT; left), the NLFFF from the magneto-frictional relaxation method (labelled NLFFF-R; middle), and the NLFFF from the optimization method (labelled NLFFF-O; right). The magnetic field lines were integrated from the same -randomly selected- photospheric footpoints for all three extrapolations (the same colour is used for the same footpoint). The gray scale displays the photospheric vertical magnetic field, $B_z$, with $\pm 500$ Gauss isocontours (purple and cyan solid lines).}
              \label{fig:Fig-B3D}
   \end{figure*} 


\subsection{Estimation of Helicity-Flux Uncertainties} \label{sec:S-Uncertainties}

As part of testing the reliability of the connectivity-based helicity flux density method, we also 
wish to evaluate uncertainties in $\gth$ and $\gph$ maps caused by those in the photospheric magnetic 
field. 
\kev{In particular, our error analysis includes both the effect of photon noise ($\approx 10$ Gauss) 
and several sources of systematic errors. \cite{Hoeksema14} recently performed an extensive analysis 
of the uncertainties in the measurements of the magnetic field strength by SDO/HMI. In particular, 
they focused on the uncertainty analysis for NOAA 11158. Among several sources of systematic errors, 
including those related with the inversion code, they found that the dominant contribution of uncertainty 
in SDO/HMI magnetic field measurements is coming from the daily variation of the radial velocity 
of the spacecraft along its geosynchronous orbit. Their analysis concludes that the typical uncertainty 
in SDO/HMI measurements of the magnetic field strength is about 100 Gauss. The latter can be easily 
checked from the estimated field-strength error map of NOAA 11158 provided by the SDO/HMI pipeline 
and which can be downloaded from JSOC\footnote{http://jsoc.stanford.edu/ajax/lookdata.html}.} 
We also want to compare these errors with those related 
with the choice of magnetic field extrapolation used to compute $\gph$.

%
%
For that purpose, we conduct a Monte Carlo experiment as proposed by \cite{Liu12} for error 
estimation of the total helicity flux. Random noise with a Gaussian distribution having a width 
($\sigma$) of 100 Gauss is added to all three components of the magnetic field for both vector 
magnetograms taken at 06:22 UT and 06:34 UT. 100 Gauss is the $\approx 1 \sigma$ uncertainty 
in the total magnetic field strength from HMI data estimated by \cite{Hoeksema14} 
\kev{and further reported in \cite{Bobra14}. The uncertainty} 
is then propagated through 
the chain of helicity flux density calculations to the flux transport velocity field 
at the photosphere derived from DAVE4VM, the corresponding vector magnetogram at $\sim$ 06:28 UT, 
and finally to the $\gth$ and $\gph$ maps.

%
%
Although they did not perform a full parametric analysis, \cite{Wiegelmann06} showed that 
extrapolations with the optimization method were not significantly affected by modest noise 
in the photospheric vector magnetogram. They found that the preprocessing of the photospheric 
data towards a more force-free state (see \eg \sect{S-NLFFF-R}) strongly helps in that matter. 
This may be expected considering that the random noise on the photospheric vector magnetic 
field acts as a source of non-force-free signals on high spatial frequencies while the preprocessing 
filters such signals out. Even if FFF reconstructions would likely be differently affected by noise 
in the photospheric data, its global effect on the extrapolations should be limited by the preprocessing 
stage. On the other hand, FFF extrapolations are more likely to be affected by more global effects 
including, but not limited to, large-scale non-force-free regions not suppressed by the preprocessing, 
magnetic flux imbalance, and lateral boundary conditions. Then, as far as $\gph$ is concerned, 
the effect of random noise on the extrapolation results is to introduce uncertainties in the connectivity 
of the magnetic field. In this regard, we believe that the choice of extrapolation method and preprocessing 
level is more fundamental and has a stronger impact on the magnetic connectivity, and hence, on $\gph$. 
For these reasons, we decided not to propagate the noise to the FFF extrapolations.

%
%
The noise propagation experiment is repeated 100 times, producing 100 noise-added $\gth$ 
and $\gph$ maps for all three FFF extrapolations considered in this paper. For each scalar quantity, 
$\mathcal{F}$, the $1\sigma$ estimated error, $\sigma_{\mathcal{F}}$, is computed as the mean, 
over all the $n$ pixels of the map, of the root mean square of the $N=100$ noise-added $\mathcal{F}$-maps, 
$\mathcal{F}_{\mathrm{n.a.}}^{i} \left( \xx_j \right) $, compared with the no-noise $\mathcal{F}$-map, 
$\mathcal{F}_{\mathrm{n.n.}} \left( \xx_j \right)$
\BE
	\label{eq:Eq-Sigma}
	\sigma_{\mathcal{F}} =  
	\frac{1}{n} \sum^{n}_{j=1} \left(
	\frac{1}{N} \sum^{N}_{i=1} 
	 \left( \mathcal{F}_{\mathrm{n.a.}}^{i} \left( \xx_j \right) 
	      - \mathcal{F}_{\mathrm{n.n.}} \left( \xx_j \right) \right)^2 
	                           \right)^{\frac{1}{2}}     \,.
\EE


\section{Force-Free Magnetic Field Extrapolations} \label{sec:S-Extrapolations}

We perform three force-free field extrapolations using different assumptions and methods: 
(1) the potential magnetic field, (2) a nonlinear FFF (NLFFF) reconstruction using 
the magneto-frictional method of \cite{Valori10}, and (3) an NLFFF generated 
with the optimization method of \cite{Wiegelmann04}. The extrapolations and setup used 
to produce them are described hereafter.

At this stage, we wish to emphasize that many more extrapolation assumptions, methods, 
and implementations, exist in the literature. 
\kev{These different methods are further distinguished in terms of the physical information 
that they extract from the photospheric vector magnetograms and use as boundary conditions. 
For instance, some NLFFF codes use vector magnetograms as boundary conditions, as is the case 
of the magneto-frictional relaxation implemented by, \eg \cite{Valori10}, or the optimization 
method implemented by, \eg \cite{Wiegelmann10}. Others built on the Grad-Rubin approach 
\citep{Grad58} use the normal component of the photospheric magnetic field and the force-free 
parameter -- derived from the vector magnetograms -- as boundary conditions 
\citep[\eg][]{Amari06,Wheatland07}. Recently, \cite{Malanushenko12} also proposed 
a Quasi-Grad-Rubin method that only uses the photospheric normal magnetic field, but combined 
with coronal loops fitting to constrain the coronal distribution of the force-free parameter. 
Finally, several codes haven been recently developed to perform a full MHD relaxation using 
photospheric vector magnetograms, thus distinguishing them from NLFFF models through the inclusion
of plasma forces in the reconstruction of the 3D magnetic field of ARs \citep[\eg][]{Inoue11,Jiang12b,Zhu13}. 
All these different codes and methods can be used to model the coronal magnetic field of ARs 
from which we can derive the magnetic connectivity required to use the connectivity-based 
helicity flux density approach to map the helicity flux in ARs.}

Several of these methods have been compared with each others in \eg \citet{DeRosa09,DeRosa15}, 
including the two NLFFF methods used in this paper. From these studies, 
it appears that the three extrapolation methods considered here generally produce differences 
in field-lines distribution that are representative of the differences that can be expected between 
the coronal magnetic fields reconstructed with other FFF methods. We thus expect the differences 
in the $\mathrm{d} h_{\Phi} / \mathrm{d} t$ and $\gph$ calculations presented in this paper to be 
representative of the differences that would be obtained when computing the connectivity-based 
helicity flux density with other extrapolation methods. For this reason, we limit ourselves 
to the analysis of helicity flux density calculations with the three extrapolations described below.

\kev{The extrapolation presented hereafter are performed on a finite field of view with open side 
and top boundaries, but without assuming magnetic flux balance. Magnetic flux is thus free to leave 
the extrapolation domain as open-like magnetic field. Such open-like magnetic field may represent 
truly open magnetic field and/or connections with the very distant quiet Sun and surrounding ARs. 
Helicity flux density calculations are not performed for open-like field lines because $\gph$ and 
$\mathrm{d} h_{\Phi} / \mathrm{d} t$ are not defined for open magnetic flux tubes (\cf \sect{S-Numerical-Method}). 
For that reason, open field lines are not plotted in any of the field-line plots presented throughout 
the paper.
	
Working with a finite field of view may be a strong limitation to reconstruct the magnetic field 
of ARs. Not only is the entire photosphere always populated with quiet Sun magnetic flux, but there are 
also often more than one AR on the Sun at a given time \citep[\eg][]{Schrijver13}. The effect of using 
a limited field of view is to remove large-scale connections with the distant quiet Sun and surrounding 
ARs that are outside the field of view considered for the extrapolation. Such distant connections may 
influence the results of the helicity flux density calculations with the connectivity-based method. 
To test such an influence, one needs to compare the connectivity-based helicity flux density computed 
from a global, full-Sun magnetic field reconstruction in spherical geometry vs. from a local magnetic 
field extrapolation. However, current global reconstructions of the solar magnetic field are limited 
by the fact that there is no full-Sun vector magnetograms at any single time, and hence, no proper 
data for the boundary conditions of full-Sun extrapolation codes. A proper analysis of the effect 
of distant ARs on the helicity flux density calculations would therefore require numerical modeling. 
Such a study is not the goal of the present work. The vast majority of investigations relying on NLFFF 
extrapolations are focusing on single ARs with the same type of limited field of view and inherent 
hypotheses that are also used in the present manuscript. We are presently testing the reliability 
of connectivity-based helicity flux density calculations with regard to such NLFFF modeling, \ie 
we aim to determine if and to which extent different choices of NLFFF computation schemes applied 
to a compact active region can impact the distribution of the helicity flux density.}


\subsection{Potential Field} \label{sec:S-POT}

The current-free magnetic field is the minimal-energy possible state for the given distribution 
of magnetic field at the boundaries of the considered volume. In addition, the potential field is 
often used as an initial state of numerical methods that build the more complex NLFFF models 
\citep[see review by, \eg][]{Wiegelmann12b}. It is therefore an important candidate to consider 
for testing the connectivity-based helicity flux density method, despite its limitation in reproducing 
nonlinear features of the coronal field.

The potential field can be directly computed using the potential theory and the reflection principle 
to solve the Laplace equation for the scalar potential in terms of the flux through the photospheric 
boundary \citep{Schmidt64}. However, in order to speed up the calculation, such a method is 
actually used only to compute the scalar potential on all six boundaries of the considered volume. 
The magnetic scalar potential in the volume is then computed solving the Laplace equation, 
subjected to the obtained Dirichlet boundary conditions, using a fast Helmholtz solver from the Intel 
Mathematical Kernel Library. For the required vertical component of the field on the lower boundary, 
the same vertical component as for the NLFFF extrapolation described \sect{S-NLFFF-R} is used. 
The potential field extrapolation is referred to as POT in the following and selected field lines 
for this extrapolation are shown in the leftmost panel of \fig{Fig-B3D}.


\subsection{NLFFF from Magneto-Frictional Relaxation} \label{sec:S-NLFFF-R}

NLFFF extrapolations are models of the coronal magnetic field that assume the corona to be static 
on the time scale of interest, and to be dominated by magnetic forces that are distributed in a way 
such that the resulting Lorentz force is everywhere vanishing. Such assumptions are supposed 
to be valid in the entire volume of interest, boundaries included. The magneto-frictional relaxation 
method implements numerical relaxation and multi-grid techniques to solve the corresponding 
equations \citep[see][for more details]{Valori07,Valori10,DeRosa15}.

The remapped and disambiguated vector magnetograms from the HMI-SHARP data series were 
interpolated to 1$''$-resolution and averaged to construct the vector magnetogram at 06:28 UT 
(\cf \sect{S-Data}) used for the NLFFF extrapolation. Vector magnetograms are inferred from 
spectropolarimetric measurements taken at photospheric heights, where the plasma is non-force-free. 
Therefore, in order to use the vector magnetogram as a boundary condition for the NLFFF extrapolation 
code, the forces acting on the magnetogram need to be reduced (preprocessing). To this purpose 
we use the method of \cite{Fuhrmann07,Fuhrmann11}. In this application, only the horizontal 
components of the field are preprocessed, yielding a reduction of the forces from 0.035 to 0.002 
in the non-dimensional units used in \cite{Metcalf08}. Since smoothing is not necessarily facilitating 
the extrapolation \citep{Valori13}, no smoothing was applied. The resulting magnetogram was then 
extrapolated using the magneto-frictional code into a volume of about $208\times202\times145$ Mm$^3$.

The resulting extrapolated field has solenoidal errors that can be quantified using the formula by 
\cite{Valori13} into $9\%$ of the total magnetic energy. The fraction of the total current that is 
perpendicular to the field is $\sigma_J=0.48$ \citep{Wheatland00}, a rather high value that is not 
uncommon for extrapolation of HMI vector magnetograms with the magneto-frictional method 
\citep[see][for an application to Hinode/SP magnetogram with a much lower $\sigma_J$]{Valori12}. 
Selected field lines of the NLFFF obtained in this way, and referred to as NLFFF-R in the following, 
are shown in the central panel of \fig{Fig-B3D}.

  \begin{figure*}
   \centerline{\includegraphics[width=0.85\textwidth,clip=]{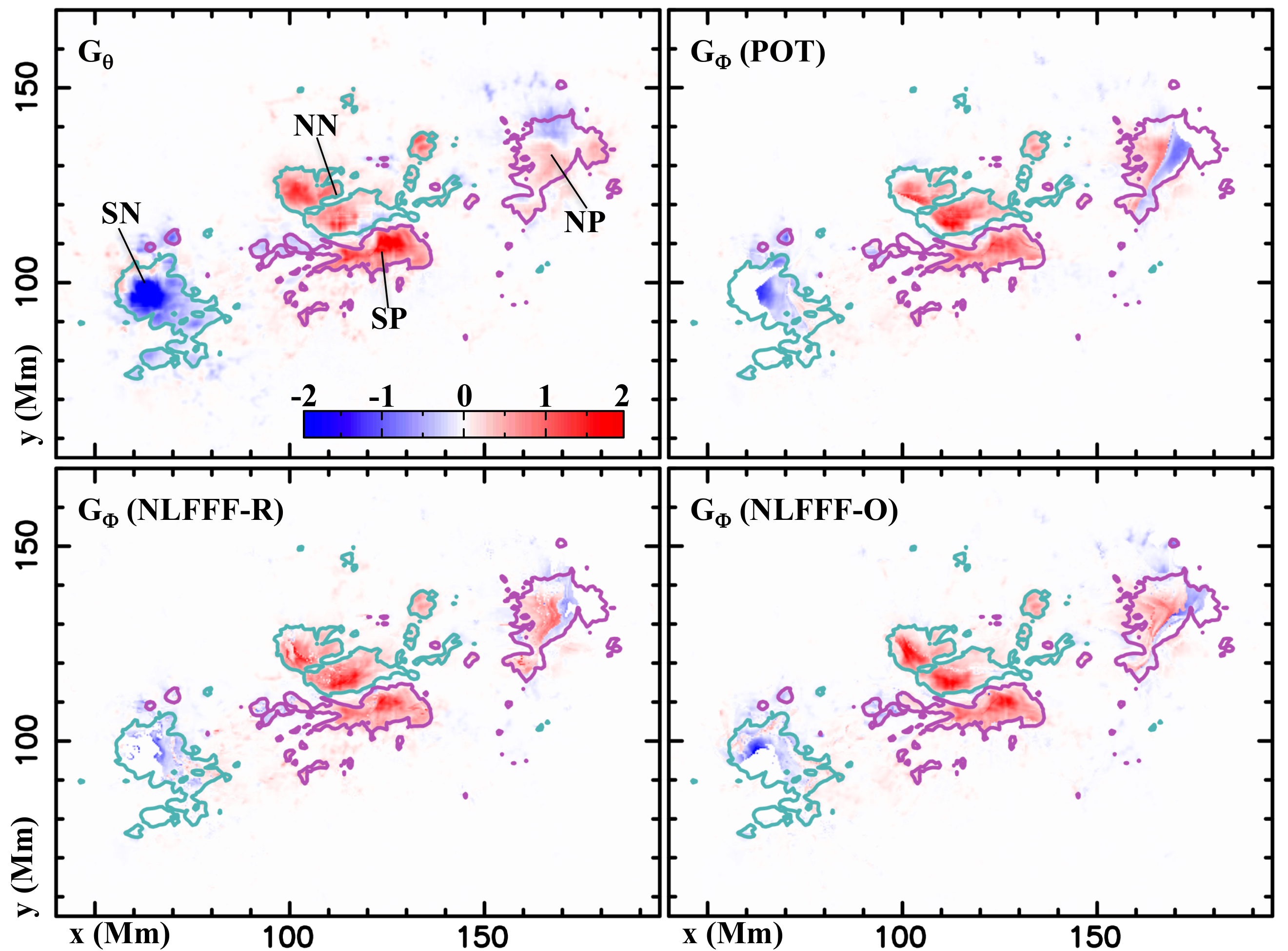}
              }
   \caption{Surface densities of helicity flux for AR 11158 at $\sim$ 06:28 UT on 2011/02/14 (in units of $10^7$ Wb$^2$ m$^{-2}$ s$^{-1}$) computed from the purely photospheric proxy ($\gth$; top left) and the connectivity-based proxy ($\gph$) derived using the three FFF extrapolations (top right, and bottom row). Purple and cyan solid lines show $B_z = \pm 500 \ \mathrm{Gauss}$ isocontours from the original photospheric vector magnetogram of SDO/HMI. The presence of strong white (\ie zero) signals in the left part of SN and north-right part of NP for all three $\gph$ maps is due to open-like magnetic flux where $\mathrm{d} h_{\Phi} / \mathrm{d} t$ and $\gph$ are not defined, and hence, not computed.}
              \label{fig:Fig-surface-densities}
   \end{figure*}


\subsection{NLFFF from Optimization Method} \label{sec:S-NLFFF-O}

For the second NLFFF model considered in this paper\gr{,} we use the weighted optimization 
method \citep{Wiegelmann04}, which is an implementation and modification of the original 
optimization algorithm of \cite{Wheatland00}. The optimization method minimizes an integrated 
joint measure, which comprises the normalized Lorentz force, the magnetic field divergence, 
and treatment of the measurement errors, over the computational domain 
\citep[see][for more details]{Wiegelmann10, Wiegelmann12}.

To perform the extrapolation, the vector magnetogram (\fig{Fig-BFTV}) is first rebinned to 1$''$ 
per pixel and preprocessed towards the force-free condition using the method of \cite{Wiegelmann06}. 
The extrapolation is finally performed on a uniform grid of $256 \times 256 \times 200$ points covering 
$\sim 185 \times 185 \times 144$ Mm$^3$. We find a solenoidal error of $1\%$ of the total magnetic 
energy and a fraction of total current perpendicular to the magnetic field $\sigma_J=0.20$. These 
values are lower than for the NLFFF-R model (\sect{S-NLFFF-R}), which is due both to different 
preprocessing and extrapolation methods and strategies.

In the following, the NLFFF model built with the optimization method is referred to as NLFFF-O. 
A set of selected magnetic field lines is shown in the rightmost panel of \fig{Fig-B3D}.


\section{Results} \label{sec:S-Results}

In this section, we analyze the results from the connectivity-based helicity flux density calculations. 
The validation of the maps, error estimations, and qualitative comparisons are briefly presented 
in \sect{S-Hflux-density}. The one-to-one quantitive comparisons are discussed in \sect{S-Quantitative-Analysis}.

\subsection{Helicity-Flux Density Distribution} \label{sec:S-Hflux-density}

   \begin{deluxetable}{c c c c c}
   \tablewidth{0.48\textwidth}
   \tablecaption{Metrics for validation and quantification of $\gph$ maps
   	\label{tab:Tab-Metrics-Gphi}
	}
   \tablehead{ \colhead{FFF model} & \colhead{$\tau_{\Phi_{\mathrm{imb.}}}$} & \colhead{$C_{\surf_{\Phi_{\mathrm{cl.}}}}$} & \colhead{$C_+$} & \colhead{$C_-$} }
   \tablecomments{The table presents the validation and quantification metrics for the maps displayed in \fig{Fig-surface-densities}. All metrics are dimensionless ratios defined by \eqss{Eq-Tau}{Eq-Cm}.}
   \startdata
	POT & \phs $1.2 \times 10^{-3}$ & $4.2 \times 10^{-4}$ & 1.32 & 1.71  \\
	NLFFF-R & $-7.6 \times 10^{-3}$ & $4.7 \times 10^{-2}$ & 1.21 & 1.67  \\
	NLFFF-O & \phs $2.6 \times 10^{-3}$ & $3.2 \times 10^{-2}$ & 1.23 & 1.57
   \enddata
   \end{deluxetable} 

%
%
\tab{Tab-Metrics-Gphi} presents the validation metrics for the $\gph$ maps computed from the three 
FFF extrapolation models described \sect{S-Extrapolations}. For all three $\gph$ maps, we obtain 
$\tau_{\Phi_{\mathrm{imb.}}}$ below $1\%$ and $C_{\surf_{\Phi_{\mathrm{cl.}}}}$ below $5\%$. 
This allows us to verify that the magnetic flux over which the connectivity-based helicity flux density is 
computed is very well balanced and that the calculation of $\mathrm{d} h_{\Phi} / \mathrm{d} t$ and 
$\gph$ well preserves the total helicity flux in that region in the closed magnetic flux. Together, 
these numbers enable us to confirm the accuracy of the $\gph$ and $\mathrm{d} h_{\Phi} / \mathrm{d} t$ 
calculations discussed in the following.

  \begin{figure*}
   \centerline{\includegraphics[width=0.98\textwidth,clip=]{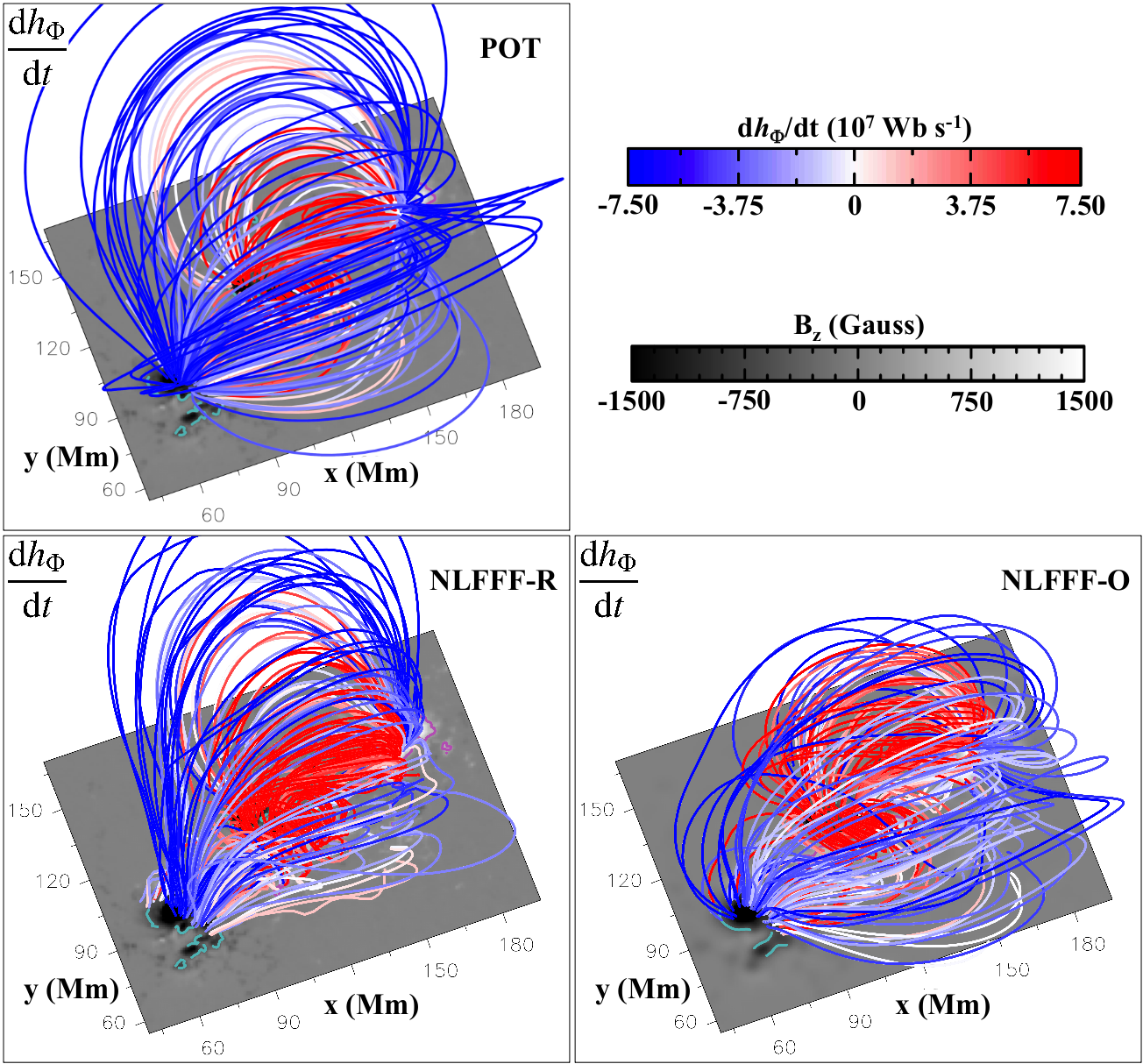}
              }
   \caption{3D representation of the connectivity-based helicity flux density for each FFF extrapolation. The magnetic field lines were integrated from the same -randomly selected- photospheric footpoints for all three extrapolations. They are colored according to their $\mathrm{d} h_{\Phi} / \mathrm{d} t$ value. Purple and cyan solid lines show $B_z = \pm 500 \ \mathrm{Gauss}$ isocontours from the FFF extrapolation.}
              \label{fig:Fig-dhphi}
   \end{figure*} 

%
%
\fig{Fig-surface-densities} presents the surface density of helicity flux from the purely photospheric 
proxy, $\gth$, and the connectivity-based proxy, $\gph$, computed from all three FFF extrapolations. 
The mean of the absolute value of the helicity flux signal in most of the AR (\ie $|B_z| \ge 100$ G) 
is $2.8 \times 10^6$ Wb$^2$ m$^{-2}$ s$^{-1}$ for the $\gth$ map and 
$1.7 \times 10^6$ Wb$^2$ m$^{-2}$ s$^{-1}$ for the $\gph$ maps, while a large fraction of the maps 
is associated with local helicity fluxes of $10^7$ Wb$^2$ m$^{-2}$ s$^{-1}$. As shown in \tab{Tab-Error-MC}, 
the errors for $\gth$ and $\gph$ estimated from our Monte Carlo experiment are 
$3.7 \times 10^5$ Wb$^2$ m$^{-2}$ s$^{-1}$ for $\gth$ and lower than 
$3 \times 10^5$ Wb$^2$ m$^{-2}$ s$^{-1}$ for $\gph$. The signal intensity of the surface density maps 
in \fig{Fig-surface-densities} is thus well above the estimated noise level.

   \begin{deluxetable}{c c c c c}
   \tablewidth{0.48\textwidth}
   \tablecaption{Error estimations from Monte Carlo experiment
   	\label{tab:Tab-Error-MC}
	}
   \tablehead{  & \colhead{$\gth$} & \colhead{$\gph($POT$)$} & \colhead{$\gph($NLFFF-R$)$} & \colhead{$\gph($NLFFF-O$)$} }
   \tablecomments{The errors are in units of $10^5$ Wb$^2$ m$^{-2}$ s$^{-1}$, \ie $\sim 10-100$ times smaller than the typical values displayed by the $\gth$ and $\gph$ maps from \fig{Fig-surface-densities}. See \sect{S-Uncertainties} for a description of error calculations.}
   \startdata
	$\sigma$ & 3.7 & 2.6 & 2.2 & 3.0
   \enddata
   \end{deluxetable} 

%
%
\fig{Fig-surface-densities} shows that the largest differences in helicity flux density maps are between 
$\gth$ and the three $\gph$ maps. In particular, the strongest differences are associated with magnetic 
flux systems that connect footpoints of opposite $\gth$ signs, \ie footpoints of NN connected to NP, 
footpoints of SN connected to SP, and footpoints of SN connected to NP. On the other hand, the $\gth$ 
map is relatively similar to the three $\gph$ maps for the flux system connecting NN to SP, because 
magnetic field lines are connecting footpoints with similar values of $\gth (\xx)/ |B_n (\xx)|$. This is 
consistent with the work of \cite{Pariat05} and \cite{Dalmasse14} who showed that $\gth$ hides 
the true helicity flux signal when simultaneous opposite helicity fluxes are present in a magnetic 
configuration. This effect is inherent to the definition of $\gth$ that does not acknowledge the fact 
that the variation of magnetic helicity in an elementary magnetic flux tube comes from the motions 
of its two photospheric footpoints with respect to the other elementary flux tubes of the entire magnetic 
configuration. As a result, the comparison of the total positive helicity flux and total negative helicity 
flux from $\gth$ in the closed magnetic flux with the same quantities computed for each one of 
the $\gph$ maps leads values of $C_+ > 1.2$ and $C_- > 1.5$ (see \tab{Tab-Metrics-Gphi}). 
Such values indicate that $\gth$ is affected by moderate spurious positive signals and rather 
high spurious negative helicity fluxes that are corrected for by the use of $\gph$ (\cf \sect{S-Metrics}). 
\fig{Fig-surface-densities} thus highlights the fact that the redistribution of helicity flux 
operated by $\gph$ is crucial to the photospheric mapping of helicity flux in ARs, regardless of 
the coronal magnetic field modeling.

All three $\gph$ maps in \fig{Fig-surface-densities} are in very good qualitative agreement, showing 
(i) a negative helicity flux in SN, (ii) a strong positive helicity flux in NN and SP, and (iii) strong positive 
and negative helicity fluxes with a strongly marked interface in NP. The $\gph$ maps from the two 
NLFFF models are very similar. Setting aside the white signal in SN, which is due to open field-lines 
where $\gph$ is not computed, and focusing on the common areas where all three extrapolations have 
closed magnetic flux, we find that the helicity flux density map derived from the potential field is also 
similar to the $\gph$ maps derived from the two NLFFF models. The most noticeable difference is 
in the south-east part of NN that does not show any significant helicity flux for $\gph($POT$)$, 
contrary to both $\gph($NLFFF-R$)$ and $\gph($NLFFF-O$)$.

Finally, the good qualitative agreement between the connectivity-based helicity flux density calculations 
from the three FFF extrapolations is further emphasized by the 3D representation of 
$\mathrm{d} h_{\Phi} / \mathrm{d} t$ in \fig{Fig-dhphi}. They all show that the inner part of the AR is 
dominated by strong positive helicity fluxes and is embedded within a magnetic field region dominated 
by strong negative helicity fluxes. The core results of \cite{Dalmasse13} are thus confirmed with a weak 
dependance on the extrapolation method.


\subsection{Quantitative Analysis} \label{sec:S-Quantitative-Analysis}

We now focus on the quantitative comparison of the helicity flux density calculations obtained 
from the three FFF models. We restrict our analysis to the pixels of the $\gph$ maps that are 
above the noise level (different for each map) estimated from the Monte Carlo experiment. 
The error levels are given in \tab{Tab-Error-MC}.

  \begin{figure}
   \centerline{\includegraphics[width=0.48\textwidth,clip=]{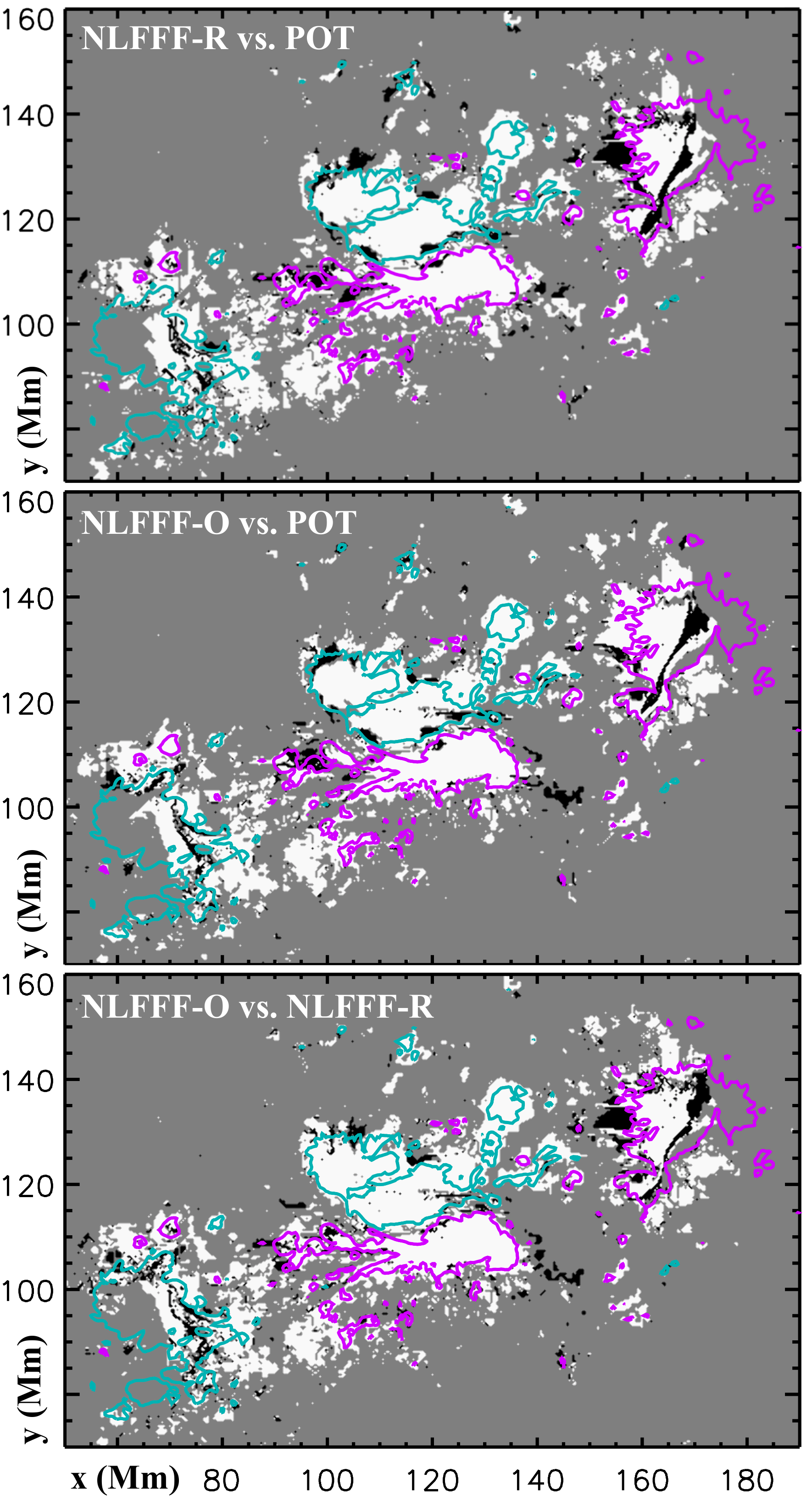}
              }
   \caption{Maps of $\gph$-sign agreement between the surface-density maps obtained with the three FFF models. For each map, $[$white; black$] = [$agree; disagree$]$, while gray corresponds to pixels that are either associated with open-like field-lines (where $\gph$ is not computed) or where $\gph$ is below the noise level for at least one of the two models being compared. Purple and cyan solid lines show $B_z = \pm 500 \ \mathrm{Gauss}$ isocontours from the original photospheric magnetogram of SDO/HMI.}
              \label{fig:Fig-Sign-Agreement}
   \end{figure}

%
%
\fig{Fig-Sign-Agreement} displays the three maps of sign agreement. The regions where our analysis 
can be carried out (white and black) is mostly associated with the strong magnetic field of AR 11158, 
\ie where $| B_z | \gtrsim 500$ Gauss. These regions correspond to the area where most of the helicity 
flux (at least $88\%$ of the total unsigned helicity flux) computed with $\gph$ is coming from. This is 
because the helicity flux intensity outside these regions is below the noise level of the respective 
$\gph$ maps. Despite the presence of some relatively small areas of disagreement (black patches), 
we find that these regions are dominated by agreement (white signal) over the sign of helicity flux 
derived using different FFF extrapolations. In particular, the percentage of surface area for which 
pairs of $\gph$ maps agree is always larger than $85 \%$, which translates into more than $\approx 84 \%$ 
in terms of magnetic flux. Therefore, the local sign of helicity flux computed from the connectivity-based 
helicity flux density method is very robust to the different FFF models and assumptions used for calculations.

  \begin{figure*}
   \centerline{\includegraphics[width=0.98\textwidth,clip=]{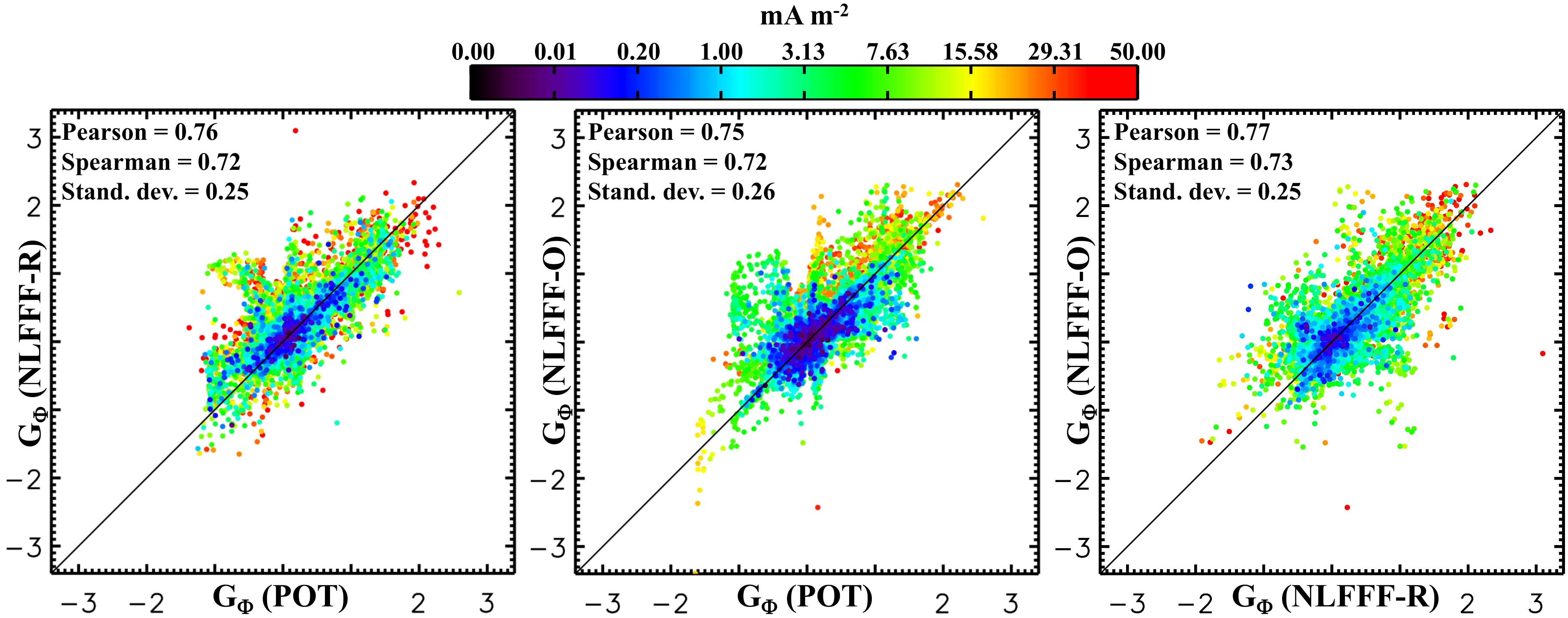}
              }
   \caption{Scatter plots of pixel-to-pixel comparison of the surface-density of helicity flux. The black solid line shows the $y=x$ line. ``Pearson'', ``Spearman'', and ``Stand. dev.'' respectively are the Pearson correlation coefficient, the Spearman correlation coefficient, and the standard deviation. The standard deviation is in units of $10^7$ Wb$^2$ m$^{-2}$ s$^{-1}$. From the left to the right panel, the color scale corresponds to $|j_z ($NLFFF-R$)|$, $|j_z ($NLFFF-O$)|$, and $ \sqrt{|j_z (\textrm{NLFFF-R})| \cdot |j_z (\textrm{NLFFF-O})| }$. Notice the color scale for the current density is not linear, but was instead chosen as $\| \cdot \|^{1/4}$ with saturation at $50$ mA m$^{-2}$ for dynamic range optimization.}
              \label{fig:Fig-Scatter}
   \end{figure*}

%
%
\fig{Fig-Scatter} displays the linear correlation plots of $\gph$ values in each pixel, for different pairs 
of FFF extrapolations. For comparisons between $\gph$ from the potential field model and one of 
the two NLFFFs, the points are colored according to the strength of the photospheric electric current 
density, $|j_z|$, from the different preprocessed boundary employed in the NLFFF model under 
consideration. For the plot comparing $\gph$ from the two NLFFF extrapolations, the points are 
colored according to $( |j_z ($NLFFF-R$)| \cdot |j_z ($NLFFF-O$)| )^{1/2}$. Such a color coding 
was introduced in order to investigate the dependency of the scatter on the electric current density 
of magnetic field lines where $\gph$ is computed.

%
%
In each scatter plot, we find that the spatial distribution of points exhibits a clear ellipsoidal shape 
aligned along the $y=x$ diagonal line. Each one of these distributions display a moderate dispersion. 
The three standard deviations computed from each scatter plot of \fig{Fig-Scatter} are $\le 2.6 \times 10^6$ 
Wb$^2$ m$^{-2}$ s$^{-1}$, which is 5 to 10 times smaller than most of the signal in the four main 
magnetic polarities. These standard deviations are $\approx 10$ times larger than the $\gph$ errors 
estimated from the Monte Carlo experiment (\tab{Tab-Error-MC}). As anticipated, it means that, 
despite the substantial agreement on the sign of the injected helicity between different extrapolations, 
a significant uncertainty in the connectivity-based calculations is coming from the choice of extrapolation 
model used to derive the field line connectivity.

   \begin{deluxetable}{c c c c c}
   \tablewidth{0.48\textwidth}
   \tablecaption{Scatter plots dispersion vs. electric current density
   	\label{tab:Tab-sigma-vs-jz}
	}
   \tablehead{\colhead{$j_z$} & \colhead{$]0, 0.2]$} & \colhead{$]0.2, 3.1]$} & \colhead{$]3.1, 15.6]$} & \colhead{$]15.6,  50.0]$} }
   \tablecomments{The interval for the standard deviation, $\sigma_{\gph}$, is derived from the values obtained for the three scatter plots of \fig{Fig-Scatter}. The electric current density is in units of mA m$^{-2}$ and the dispersion in units of $10^6$ Wb$^2$ m$^{-2}$ s$^{-1}$. From left to right, the five ranges of electric current density correspond to $]\mathrm{black}; \ \mathrm{dark \ blue}]$, $]\mathrm{dark \ blue}; \ \mathrm{green}]$, $]\mathrm{green}; \ \mathrm{yellow}]$, and $]\mathrm{yellow}; \ \mathrm{red}]$ of the color-scale in \fig{Fig-Scatter}.}
   \startdata
	 \phs $\sigma_{\gph}$ & $[1.5, 1.8]$ & $[1.8, 2.0]$ & $[2.6, 3.6]$ & $[3.0, 4.1]$
   \enddata
   \end{deluxetable} 

%
%
\fig{Fig-Scatter} further shows that the ellipsoidal pattern of the distribution of points is present 
independently of the electric current density, \ie of the color of the points. We also notice that 
the scattering of points away from the $y=x$ line presents some relatively weak dependance 
on the electric current density of the field lines used for the connectivity-based calculations. 
In particular, we find a dispersion in the range $\approx [1.5, 2.0] \times 10^6$ Wb$^2$ m$^{-2}$ s$^{-1}$ 
for black to green points, and $\approx [2.6, 4.1] \times 10^6$ Wb$^2$ m$^{-2}$ s$^{-1}$ for green 
to red points (finer details are provided in \tab{Tab-sigma-vs-jz}). In addition, the number 
of pixels with average-to-strong electric current density (\ie green to red points) is sensibly 
the same as the number of pixels with very weak-to-average electric current density (\ie black 
to green points). This implies that the correlation coefficients, displayed in \fig{Fig-Scatter} and 
discussed below, are not dominated by the differences in $\gph$ values from nearly-potential 
magnetic field lines.

%
%
For all three scatter plots, we find that the Pearson, $c_P$, and Spearman, $c_S$, correlation 
coefficients are such that $c_P \ge 0.75$ and $c_S \ge 0.72$. We checked that these values are 
statistically significant by conducting a {\it null hypothesis} test (details and results of this test are 
provided in \app{A-Null-hypothesis}). We therefore conclude that the calculations of the connectivity-based 
helicity flux density derived from different FFF extrapolations are highly correlated and consistent 
with each other.

%
%
\kev{For further comparison, we compute the vector correlation metric, $C_{\mathrm{vec}}$, 
comparing the three 3D magnetic field extrapolations as defined by Equation (28) of \cite{Schrijver06}. 
For that purpose, the POT and NLFFF-R 3D magnetic fields are interpolated on the same grid 
as the NLFFF-O (whose extrapolation domain is common to all three models) using trilinear 
interpolation. We find $C_{\mathrm{vec}} (\bb_{\mathrm{NLFFF-R}},\bb_{\mathrm{POT}}) = 0.84$, 
$C_{\mathrm{vec}} (\bb_{\mathrm{NLFFF-O}},\bb_{\mathrm{POT}}) = 0.90$, 
and $C_{\mathrm{vec}} (\bb_{\mathrm{NLFFF-O}},\bb_{\mathrm{NLFFF-R}}) = 0.90$. Such values for the vector 
correlation metric of the magnetic fields are very high even though the 3D distributions of the magnetic 
field lines are relatively different when comparing the three extrapolations, as inferred from \fig{Fig-B3D}. 
We thus find that the vector correlation metric for the magnetic fields is higher than the Pearson and 
Spearman correlation coefficients found when comparing the helicity flux density calculations.}


\section{Discussion} \label{sec:S-Discussion}

  \begin{figure}
   \centerline{\includegraphics[width=0.48\textwidth,clip=]{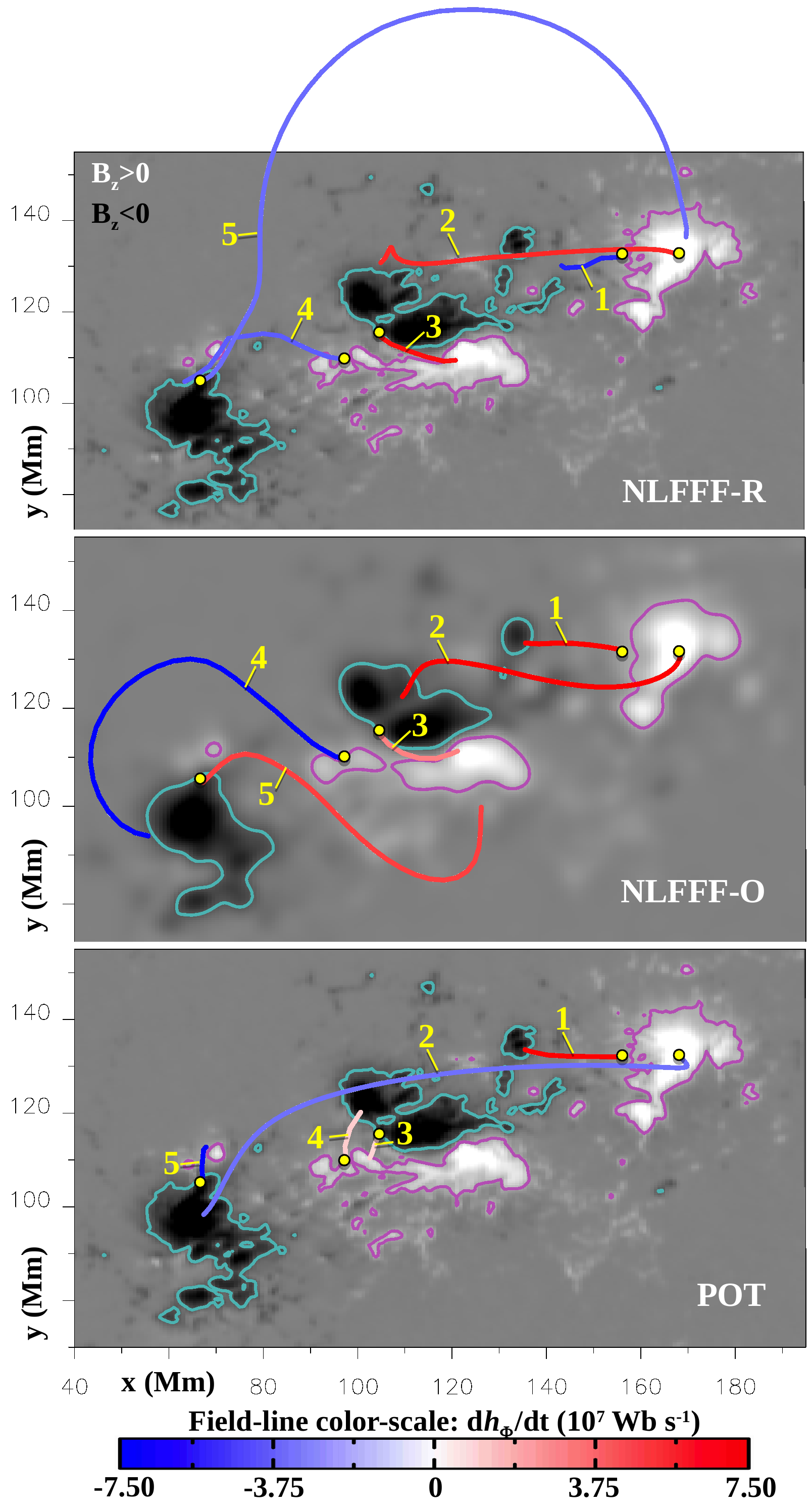}
              }
   \caption{3D representation (top view) of the connectivity-based helicity flux density, as in \fig{Fig-dhphi}, illustrating local differences between $\mathrm{d} h_{\Phi} / \mathrm{d} t$ from the different FFF extrapolations. The magnetic field lines are labelled according to the common photospheric footpoint from which they were integrated and which is indicated by the yellow disk. The color scale for the vertical magnetic field (gray scale) and its isocontours are the same as in \fig{Fig-dhphi}.}
              \label{fig:Fig-dhphi-diffs}
   \end{figure}

%
%
At this point, we wish to emphasize that the robustness of the connectivity-based method against 
different magnetic field extrapolations {\it does not} mean that the extrapolations are very much alike. 
On the contrary, all three extrapolations considered here produce 3D magnetic fields that are different 
from each other as shown in \fig{Fig-B3D}. So, what does it mean that the connectivity-based helicity 
flux density calculations are robust against different extrapolation methods?

%
%
First of all, we remind the reader that the surface-density of helicity flux, $\gph$, only explicitly 
depends on the connectivity of magnetic field lines and not on their 3D geometry; the latter only has 
an implicit effect on $\gph$ by affecting the magnetic connectivity. Secondly, expanding \eq{Eq-Gphi} 
using \eq{Eq-hflux-dens-2} leads to
\BE
	\label{eq:Eq-Gphi-expanded}
	\gph (\xx_{+}) = \frac{|B_n (\xx_{+})|}{2} \left( \dot{\Theta}_B (\xx_{+}) - \dot{\Theta}_B (\xx_{-})  \right)   \,.
\EE
\eq{Eq-Gphi-expanded} shows that, for a given footpoint $\xx_{+}$, the helicity flux density $\gph$ 
from different extrapolation models will be exactly the same (1) if they have the exact same magnetic 
field connectivity, or (2) if they have a different connectivity, the first extrapolation links $\xx_{+}$ 
to $\xx_{1_-}$ and the second links $\xx_{+}$ to $\xx_{2_-} \ne \xx_{1_-}$, but 
$\dot{\Theta}_B (\xx_{2_-}) = \dot{\Theta}_B (\xx_{1_-})$. The same type of conclusion can be drawn 
for $\gph (\xx_{-})$ by simply exchanging $\xx_{+}$ and $\xx_{-}$. Note that condition (2) relates 
to the spatial smoothness of $\dot{\Theta}_B (\xx)$ and, by extension, $\gth (\xx)$.

In general, $\gph$ can distinguish two magnetic fields that have different photospheric magnetic 
connectivities. This is evident from the connectivity-based flux density maps shown \fig{Fig-surface-densities} 
and the clear dispersion shown in the scatter plots of \fig{Fig-Scatter}. However, we argue that 
as long as two different 3D magnetic fields have, {\it on average}, a similar magnetic connectivity, then 
the $\gph$ maps computed from these magnetic fields should display a good {\it overall} agreement. 
Note that, although not discussed, this was already observed by \cite{Dalmasse14} for the models 
analyzed in their Figures 8, 10, 11, and 12. The strongest differences would then be expected in 
localized regions where the magnetic connectivity from the extrapolations is markedly different, 
typically in the close surroundings of quasi-separatrix layers (QSLs) which are regions of strong 
gradients of the magnetic connectivity that are favorable to magnetic reconnection 
\citep[\eg][]{Demoulin96,Titov02,Aulanier06,Janvier13}. This is indeed what we find in our analysis 
of AR 11158, \eg at the interface of positive and negative helicity fluxes in NP (see \fig{Fig-surface-densities}) 
that coincides with a large-scale QSL that separates field-lines connecting NP to NN and NP to SN.

%
%
Then, we recall that $\gph$ is only a 2D representation of the physical, 3D, definition of local magnetic 
helicity variation, $\mathrm{d} h_{\Phi} / \mathrm{d} t$. $\mathrm{d} h_{\Phi} / \mathrm{d} t$ is only 
defined for elementary magnetic flux tubes. Its 3D representation thus requires to plot individual 
field lines. Since different magnetic field extrapolation methods and assumptions generally produce 
3D magnetic fields that can differ significantly in the details of individual field lines \citep[\eg][]{DeRosa09}, 
then, differently from $\gph$, $\mathrm{d} h_{\Phi} / \mathrm{d} t$ can always differentiate two magnetic 
fields. This is indeed what we see in \fig{Fig-dhphi} where the three plots of $\mathrm{d} h_{\Phi} / \mathrm{d} t$ 
are easily distinguishable because of the different 3D geometry of magnetic field lines. Thus, the robustness 
of $\mathrm{d} h_{\Phi} / \mathrm{d} t$ against different extrapolation models should be understood 
in terms of average or global distribution of helicity flux density over the different magnetic flux systems 
of an AR, and not in terms of a one-to-one field-line and $\mathrm{d} h_{\Phi} / \mathrm{d} t$ 
correspondence. For AR 11158, this means looking at the AR in terms of the four flux systems NN-NP, 
NN-SP, SN-SP, and SN-NP. While the actual distribution of the magnetic field-lines and 
$\mathrm{d} h_{\Phi} / \mathrm{d} t$ in these four flux systems vary from one extrapolation to the other 
(see \fig{Fig-dhphi}), the sign and average helicity flux intensities agree very well, hence the robustness 
of the calculations.

%
%
On the other hand, local differences exist in the connectivity-based helicity flux density calculations 
performed with different extrapolations. Such local differences can be significant and are extremely 
important for physically interpretating the local helicity flux, \ie at the scale of a particular 
field line. This is illustrated in \fig{Fig-dhphi-diffs} that shows $\mathrm{d} h_{\Phi} / \mathrm{d} t$ 
for five field-lines that have been integrated from the same photospheric starting footpoints for all 
three extrapolations of AR 11158. The connectivity and 3D geometry of these field lines strongly 
differ from one extrapolation to the other, which results in different helicity flux intensities and signs. 
For instance, field-line 5 links SN to a small-scale positive magnetic polarity on the north of SN 
with a strong negative helicity flux for POT, while it links SN to NP with a medium negative helicity 
flux for NLFFF-R, and links SN to SP with a medium positive helicity flux for NLFFF-O. When 
comparing the three extrapolations, the five field-lines displayed in \fig{Fig-dhphi-diffs} are so 
different in geometry and orientation with respect to each other that the physical interpretation 
of their helicity flux density -- based on field-lines reorientation in response to the motions of 
their photospheric footpoints \citep[\cf Section 5 and Figure 9 of][]{Dalmasse14}, is entirely 
extrapolation dependent.

%
%
Considering the current limitations of magnetic field extrapolations, we conclude that the physical 
interpretation of the connectivity-based helicity flux density calculations in observational analyses 
will be robust at the scale of the different flux systems forming an AR, but not necessarily 
at the extremely local scale of individual magnetic field lines for which interpretation of the signal 
should be taken with a lot of caution.

%
%
\kev{Finally, we recall that the analysis presented in this paper was conducted using NLFFF 
extrapolations performed with a finite field of view. As mentioned \sect{S-POT}, this is 
a limitation since it disregards the effect of the distant quiet Sun and surrounding ARs 
that are outside the field of view considered for the extrapolation. Such effects would 
need another study with large-scale magnetic field extrapolations, or even with full-Sun 
numerical simulations.}


\section{Conclusion} \label{sec:S-Conclusion}

%
%
Thanks to the conservation properties of magnetic helicity in the solar atmosphere, studying 
the photospheric flux of magnetic helicity appears to be a key element for improving our understanding 
of how this fundamental quantity affects the dynamics of solar active regions (ARs). For that purpose, 
a connectivity-based helicity flux density method, built upon the work of \cite{Pariat05}, was recently 
developed and tested on various analytical case-studies and numerical magnetohydrodynamics (MHD) 
simulations \citep{Dalmasse14}. The ability of this method to correctly capture the local transfer of 
magnetic helicity relies on its exploitation of the connectivity of magnetic field lines, which enables 
it to embrace the 3D and global nature of magnetic helicity. 

%
%
For the solar atmosphere, the application of the connectivity-based helicity flux density method relies 
on approximate 3D solutions obtained from force-free field (FFF) extrapolations of the photospheric 
magnetic field to derive the connectivity of magnetic field lines. In general, such FFF models provide 
reconstructed magnetic fields whose 3D distribution strongly depends on the extrapolation method 
used \citep[\eg][]{DeRosa09,DeRosa15}. As a consequence, the values of subsequently derived 
quantities, such as free magnetic energy and magnetic helicity, exhibit large variations from one FFF 
model to another. Since the magnetic connectivity also depends on the 3D distribution of the extrapolated 
magnetic field, the connectivity-based helicity flux density calculations may be strongly affected 
by the choice of FFF reconstruction method. In this paper, we addressed this concern by applying 
the connectivity-based approach to solar observations with different magnetic field extrapolation 
models and implementations.

%
%
To assess the reliability and relevance of the connectivity-based helicity flux density method to solar 
observations, we considered the 
\kev{internally complex (several bipoles) and externally simple (i.e., no neighboring large-flux system)} 
AR 11158 using the vector magnetogram data from 
SDO/HMI. Three FFF extrapolations, \ie a potential field, a nonlinear FFF (NLFFF) extrapolation using 
the magneto-frictional method of \cite{Valori10}, and a second NLFFF from the optimization method 
of \cite{Wiegelmann04}, were performed to reconstruct the coronal magnetic field of AR 11158 and 
apply the connectivity-based approach. Our analysis indicates that the helicity flux density calculations 
derived from different FFF extrapolations are highly correlated (with Pearson and Spearman correlation 
coefficients larger than $\approx 0.72$) and consistent with each other, showing a very good agreement 
over identifying the local sign of helicity flux (\ie for more than $\approx 85 \%$ of the surface where 
they were compared). We thus conclude that the connectivity-based helicity flux density method can be 
reliably used in observational analyses of ARs.

%
%
The results presented in this paper also enable us to propose a procedure for estimating uncertainties 
in the connectivity-based helicity flux density calculations applied to solar observations, as follows:
\begin{enumerate}
	\item Perform a Monte Carlo experiment, as described in \sect{S-Uncertainties} and proposed 
		by \cite{Liu12}, by adding random noise with a Gaussian distribution to the photospheric 
		vector magnetic fields used for computation and propagate it through the chain of helicity 
		flux density calculations. This allows to estimate uncertainties related with magnetic field 
		measurement errors for the flux transport velocity, $\gth$, and $\mathrm{d} h_{\Phi} / \mathrm{d} t$ 
		and $\gph$ from the NLFFF method chosen for the analysis.
	\item Apply the connectivity-based calculations with the NLFFF and the potential field to derive 
		the standard error from the comparison of $\gph$ computed with each magnetic field model. 
		This allows to derive the error contribution related with the uncertainty in the magnetic field 
		connectivity due to the choice of magnetic field extrapolation method.
	\item Sum the squared errors of $\gph$ to estimate the overall uncertainty in helicity flux density 
		calculations.
\end{enumerate}
This procedure, and in particular step 2, is motivated by the fact that our analysis indicates that 
comparing the calculations with the potential field and one of the two NLFFF models gives a standard 
error that is extremely close to the standard error obtained from comparing the helicity flux density 
calculations from the two NLFFF models. Computing the potential field is relatively inexpensive 
as compared with computing an NLFFF model and, in fact, is already part of most NLFFF algorithms 
\citep[see \eg review by][]{Wiegelmann12b} that use it as an initial state.

%
%
The reliability of the connectivity-based helicity flux density calculations against different FFF models 
offers several interesting perspectives for analyzing the 2D and 3D transfer of magnetic helicity in solar 
ARs. During the early stages of AR formation, the connectivity-based method provides information 
on the distribution of magnetic helicity in the emerging magnetic field, which is an important constraint 
for models of generation and transport of magnetic flux in the solar convection zone 
\citep[\eg][]{Berger00,Pariat07,Vemareddy17}. 

%
%
On the other hand, the study of the helicity flux distribution at later stages of ARs evolution allows 
to track the sites where magnetic helicity is transferred to the corona, probing in this way the relationships 
between magnetic helicity accumulation and the energetics of solar flares and coronal mass ejections 
(CMEs). The connectivity-based approach may further be used to test the very high-energy flare model 
of \cite{Kusano04} based on magnetic helicity annihilation. Such a flare model requires the prior transfer 
and accumulation of magnetic helicity of opposite signs in different magnetic flux systems of an AR that 
would later reconnect together. Identifying AR candidates for hosting such a flare model requires reliable 
maps that are not polluted by false helicity flux signals of opposite signs. The present study shows that 
the connectivity-based helicity flux density method is very well suited for that purpose.

%
%
In summary, the connectivity-based helicity flux density method is a very promising tool for helping 
us unveil the role of magnetic helicity in the dynamics of the solar corona.


\begin{acknowledgements}
\kev{We thank the anonymous referee for helpful comments that improved the paper.}  
K.D. thanks D. Nychka for his help on the Monte Carlo experiment, and R. Centeno Elliott 
and A. Gri\~n\'on Mar\'in for their insights on HMI noise characterization. 
K.D. acknowledges funding from the Advanced Study Program, the High Altitude Observatory, 
and the Computational and Information Systems Laboratory. 
E.P. acknowledges the support of the French Agence Nationale pour la Recherche 
through the HELISOL project, contract n$^\circ$ ANR-15-CE31-0001. 
G.V. acknowledges the support of the Leverhulme Trust Research Project Grant 
2014-051. 
J.J. is supported by NASA grant NNX16AF72G and 80NSSC17K0016. 
The data used here are courtesy of the NASA/SDO and the HMI science team. 
This work used the DAVE4VM code written and developed by P. Schuck at the Naval 
Research Laboratory. 
The helicity flux calculations were performed partly on the high-performance computing system 
Yellowstone (ark:/85065/d7wd3xhc) provided by NCAR's Computational and Information 
Systems Laboratory, sponsored by the National Science Foundation, and partly on the multi-processors 
TRU64 computer of the LESIA. 
The National Center for Atmospheric Research is sponsored by the National Science 
Foundation.
\end{acknowledgements}

\bibliographystyle{aa}
      
\bibliography{HfluxRobustness}  

\IfFileExists{\jobname.bbl}{} {\typeout{}
\typeout{****************************************************}
\typeout{****************************************************}
\typeout{** Please run "bibtex \jobname" to obtain} \typeout{**
the bibliography and then re-run LaTeX} \typeout{** twice to fix
the references !}
\typeout{****************************************************}
\typeout{****************************************************}
\typeout{}}

\appendix

\section{Null hypothesis testing} \label{app:A-Null-hypothesis}

To determine whether the values of correlation coefficients reported in \sect{S-Quantitative-Analysis} 
are statistically significant, we perform a {\it null hypothesis} test \citep[\eg][]{Neyman33,Moore03,Cox06}. 
The null hypothesis states that an observed result, or relationship between two variables, is due 
to random processes alone. The null hypothesis test is an {\it argumentum ad absurdum} approach. 
The goal is to show that an observed relationship or result is very unlikely to occur under the null 
hypothesis, in which case the null hypothesis can be rejected and the alternative accepted. 
In the context of this paper, they can be formulated as follows
\begin{enumerate}[-]
	\item Null hypothesis: {\it the values of $\gph$ computed from different FFF models are not correlated}. 
	\item Alternative hypothesis: {\it the values of $\gph$ computed from different FFF models are correlated}.
\end{enumerate}

%
%
To test this null hypothesis, we perform a permutation test. Let $X = \{x_1, x_2, ..., x_n\}$ and 
$Y = \{y_1, y_2, ..., y_n\}$ be two datasets. $c^{(0)}_S = c_S (X,Y)$ is the Spearman correlation 
coefficient of the two original datasets $X$ and $Y$, and for which we want to determine 
the statistical significance. The permutation test consists in the following steps
\begin{enumerate}
	\item Create a new dataset $Y^{(k)}$ by randomly permuting the elements of $Y$; for instance, 
		$Y^{(k)} = \{y_4,y_{n-10}, ..., y_2\}$.
	\item Compute the Spearman correlation coefficient, $c^{(k)}_S = c_S (X,Y^{(k)})$.
	\item Repeat steps (1) and (2) $N$ times, where $N$ is large (typically larger than 1000). 
		This leads to $N$ sets of random permutations of $Y$, and hence, $N+1$ Spearman 
		correlation coefficients, $\{ c^{(0)}_S, c^{(1)}_S, c^{(2)}_S, c^{(3)}_S, ..., c^{(N)}_S \}$.
\end{enumerate}
We then determine the p-value of $c^{(0)}_S$, \ie the probability of obtaining the Spearman correlation 
coefficient, $c^{(0)}_S$, between the two original datasets, $X$ and $Y$, if the null hypothesis were 
true. The p-value associated with $c^{(0)}_S$ and estimated from the permutation test is the fraction 
of $c^{(k=\{0,1,...,n\})}_S$ that are larger than the Spearman correlation coefficient from the two original 
datasets, $c^{(0)}_S$, \ie
\BE
	\label{eq:Eq-pvalue}
	p = \frac{m}{N + 1}  \,,
\EE
where $m$ is the number of $c^{(k=\{0,1,...,n\})}_S$ that are $\ge c^{(0)}_S$. The same method is 
applied with the Pearson correlation coefficient. We reject the null hypothesis and accept the alternative 
if the estimated p-value for both the Pearson and Spearman correlation coefficients is strictly smaller 
than $0.001$, \ie the level below which we consider the correlation coefficients from the two original 
dataset to be statistically significant.

   \begin{deluxetable}{c c c c c}
   \tablewidth{0.6\textwidth}
   \tablecaption{Permutation tests for the Spearman correlation coefficient of $\gph($POT$)$ vs. $ \gph($NLFFF-R$)$
   	\label{tab:Tab-Permutation-tests}
	}
   \tablehead{
\colhead{$N$} & \colhead{$\mu_S$} & \colhead{$\sigma_S$} & \colhead{$c^{(0)}_{S}$} & \colhead{$p \left(c^{(0)}_{S} \right)$}
   }
   \tablecomments{$\mu_S$ and $\sigma_S$ are the mean and standard deviation of $c^{(k=\{0,1,...,n\})}_S$. $c^{(0)}_{S}$ is the Spearman correlation coefficient computed from the original pair of datasets and $p \left(c^{(0)}_S \right)$ its estimated p-value given the null hypothesis.}
   \startdata
	 $10^4$ & \phs $8.8 \times 10^{-5}$ & $1.0 \times 10^{-2}$ & 0.72 & $1.0 \times 10^{-4}$ \\
	 $10^5$ & $-6.3 \times 10^{-6}$ & $7.4 \times 10^{-3}$ & 0.72 & $1.0 \times 10^{-5}$ \\
	 $10^6$ & $-5.4 \times 10^{-6}$ & $7.1 \times 10^{-3}$ & 0.72 & $1.0 \times 10^{-6}$
   \enddata
   \end{deluxetable} 

%
%
The results from the permutation tests are summarized in \tab{Tab-Permutation-tests} 
for the Spearman correlation coefficient of $\gph($POT$)$ vs. $ \gph($NLFFF-R$)$ only, 
because we obtained very similar results for both the Pearson and Spearman correlation 
coefficients of all three scatter plots from \fig{Fig-Scatter}. For all permutation tests reported 
in \tab{Tab-Permutation-tests}, the distribution of Spearman correlation coefficients (not shown 
here), $c^{(k=\{0,1,...,n\})}_S$, exhibits a Gaussian-like profile with a mean, $\mu \approx 0$ 
($|\mu| < 10^{-4}$), and a very small standard deviation, $\sigma \le 10^{-2}$. We varied 
the number of random permutations and verified that the mean and standard deviation of 
the resulting distributions are not strongly dependent on the number of permutations as long 
as this number is large enough (typically $N \ge 10^4$). \tab{Tab-Permutation-tests} shows 
that the p-value of $c^{(0)}_S$ is at most $10^{-6}$ (as taken from the test with $N = 10^6$) 
for all permutation tests. This is only an upper bound for the p-value as suggested by their 
$N^{-1}$ dependency visible in \tab{Tab-Permutation-tests} and the comparison between 
$c^{(0)}_S$ and the standard deviation which places $c^{(0)}_S$ at an $\approx 100 \sigma$ 
distance from the mean in the tail of the distribution of $c^{(k=\{0,1,...,n\})}_S$. Note that the $N^{-1}$ 
dependency occurs because we obtain $c^{(k)}_S < c^{(0)}_S$ for all $k > 0$ for all permutation 
tests, leading to $m=1$ in \eq{Eq-pvalue}. The correlation coefficients reported in \fig{Fig-Scatter} 
are thus statistically significant. We thus reject the null hypothesis and accept the alternative 
hypothesis. We therefore conclude that the calculations of the connectivity-based helicity flux 
density derived from different FFF extrapolations are highly correlated and consistent with each 
other.

\end{document}